\PassOptionsToPackage{unicode}{hyperref}
\PassOptionsToPackage{hyphens}{url}
\PassOptionsToPackage{dvipsnames,svgnames,x11names}{xcolor}
\documentclass[
  11pt]{article}

\usepackage{amsmath,amssymb,amsthm,comment}
\usepackage{iftex}
\ifPDFTeX
  \usepackage[T1]{fontenc}
  \usepackage[utf8]{inputenc}
  \usepackage{textcomp} 
\else 
  \usepackage{unicode-math}
  \defaultfontfeatures{Scale=MatchLowercase}
  \defaultfontfeatures[\rmfamily]{Ligatures=TeX,Scale=1}
\fi
\usepackage{lmodern}
\ifPDFTeX\else  
\fi
\IfFileExists{upquote.sty}{\usepackage{upquote}}{}
\IfFileExists{microtype.sty}{
  \usepackage[]{microtype}
  \UseMicrotypeSet[protrusion]{basicmath} 
}{}
\makeatletter
\@ifundefined{KOMAClassName}{
  \IfFileExists{parskip.sty}{%
    \usepackage{parskip}
  }{
    \setlength{\parindent}{0pt}
    \setlength{\parskip}{6pt plus 2pt minus 1pt}}
}{
  \KOMAoptions{parskip=half}}
\makeatother
\usepackage{xcolor}
\makeatletter
\ifx\paragraph\undefined\else
  \let\oldparagraph\paragraph
  \renewcommand{\paragraph}{
    \@ifstar
      \xxxParagraphStar
      \xxxParagraphNoStar
  }
  \newcommand{\xxxParagraphStar}[1]{\oldparagraph*{#1}\mbox{}}
  \newcommand{\xxxParagraphNoStar}[1]{\oldparagraph{#1}\mbox{}}
\fi
\ifx\subparagraph\undefined\else
  \let\oldsubparagraph\subparagraph
  \renewcommand{\subparagraph}{
    \@ifstar
      \xxxSubParagraphStar
      \xxxSubParagraphNoStar
  }
  \newcommand{\xxxSubParagraphStar}[1]{\oldsubparagraph*{#1}\mbox{}}
  \newcommand{\xxxSubParagraphNoStar}[1]{\oldsubparagraph{#1}\mbox{}}
\fi
\makeatother

\usepackage{longtable,booktabs,array}
\usepackage{calc} 
\usepackage{etoolbox}
\makeatletter
\patchcmd\longtable{\par}{\if@noskipsec\mbox{}\fi\par}{}{}
\makeatother
\IfFileExists{footnotehyper.sty}{\usepackage{footnotehyper}}{\usepackage{footnote}}
\makesavenoteenv{longtable}
\usepackage{graphicx}
\makeatletter
\def\maxwidth{\ifdim\Gin@nat@width>\linewidth\linewidth\else\Gin@nat@width\fi}
\def\maxheight{\ifdim\Gin@nat@height>\textheight\textheight\else\Gin@nat@height\fi}
\makeatother
\setkeys{Gin}{width=\maxwidth,height=\maxheight,keepaspectratio}
\makeatletter
\def\fps@figure{htbp}
\makeatother

\makeatletter
\@ifpackageloaded{caption}{}{\usepackage{caption}}
\AtBeginDocument{%
\ifdefined\contentsname
  \renewcommand*\contentsname{Table of contents}
\else
  \newcommand\contentsname{Table of contents}
\fi
\ifdefined\listfigurename
  \renewcommand*\listfigurename{List of Figures}
\else
  \newcommand\listfigurename{List of Figures}
\fi
\ifdefined\listtablename
  \renewcommand*\listtablename{List of Tables}
\else
  \newcommand\listtablename{List of Tables}
\fi
\ifdefined\figurename
  \renewcommand*\figurename{Figure}
\else
  \newcommand\figurename{Figure}
\fi
\ifdefined\tablename
  \renewcommand*\tablename{Table}
\else
  \newcommand\tablename{Table}
\fi
}
\@ifpackageloaded{float}{}{\usepackage{float}}
\floatstyle{ruled}
\@ifundefined{c@chapter}{\newfloat{codelisting}{h}{lop}}{\newfloat{codelisting}{h}{lop}[chapter]}
\floatname{codelisting}{Listing}

\makeatother
\makeatletter
\@ifpackageloaded{caption}{}{\usepackage{caption}}
\@ifpackageloaded{subcaption}{}{\usepackage{subcaption}}
\makeatother

\ifLuaTeX
  \usepackage{selnolig}  
\fi
\usepackage[]{natbib}
\usepackage{bookmark}

\IfFileExists{xurl.sty}{\usepackage{xurl}}{} 
\hypersetup{
  pdftitle={Parametric Multi-Fidelity Monte Carlo Estimation\\
With Applications to Extremes},
  pdfauthor={M. Kim, B. Brown, and V. Pipiras},
  pdfkeywords={uncertainty quantification, semi-supervised learning, extreme value theory and distributions, maximum likelihood, ship motions.},
  colorlinks=true,
  linkcolor={blue},
  filecolor={Maroon},
  citecolor={Blue},
  urlcolor={Blue},
  pdfcreator={LaTeX via pandoc}}

\newcommand{\anon}{1}

\input{command}

\begin{document}

\def\spacingset#1{\renewcommand{\baselinestretch}%
{#1}\small\normalsize} \spacingset{1}


\if1\anon
{
  \title{\bf Parametric Multi-Fidelity Monte Carlo Estimation
With Applications to Extremes}
  \author{Minji Kim, Brendan Brown, and Vladas Pipiras
    \thanks{The authors gratefully acknowledge the partial support of ONR grants N00014-19-1-2092 and N00014-23-1-2176 under Dr.\ Woei-Min Lin and Dr.\ Robert Brizzolara. The authors also thank the two anonymous Reviewers, Associate Editor and Editor for their many useful comments that helped improve the paper.}\\
    Statistics and Operations Research, University of North Carolina at Chapel Hill\\}
    \date{}
  \maketitle
} \fi
\if0\anon
{
  \bigskip
  \bigskip
  \bigskip
  \begin{center}
    {\LARGE\bf Parametric multi-fidelity Monte Carlo estimation
with applications to extremes}
\end{center}
  \medskip
} \fi

\bigskip
\begin{abstract}
In a multi-fidelity setting, data are available from two sources, high- and low-fidelity. Low-fidelity data has larger size and can be leveraged to make more efficient inference about quantities of interest, e.g.\ the mean, for high-fidelity variables. In this work, such multi-fidelity setting is studied when the goal is to fit more efficiently a parametric model to high-fidelity data. Three multi-fidelity parameter estimation methods are considered, joint maximum likelihood, (multi-fidelity) moment estimation and (multi-fidelity) marginal maximum likelihood, and are illustrated on several parametric models, with the focus on parametric families used in extreme value analysis. An application is also provided concerning quantification of occurrences of extreme ship motions generated by two computer codes of varying fidelity.
\end{abstract}

\noindent%
\vfill

\newpage
\spacingset{1.8} 

\section{Introduction}
\label{s:introduction}

\subsection{Multi-fidelity setting}
In a typical multi-fidelity (MF) setting, one assumes that data come from multiple sources of varying fidelities and computational costs.
The focus of this work is on the case of two sources, a high-fidelity source and a low-fidelity one, though the general case is also touched upon. Given some (random) conditions $x$, consider thus possibly dependent random variables $\yhifi = \yhifi(x)$ and $\ylofi = \ylofi(x)$ associated with the high- and low- fidelity sources, respectively. Suppose $(\yhifi_1, \ylofi_1), \dots, (\yhifi_n, \ylofi_n)$ are i.i.d.\ observations of $(\yhifi, \ylofi)$ obtained from $n\ge 1$ repeated experiments.
A computational cost to produce $\ylofi$ is supposed to be considerably smaller (compared to $\yhifi$), at the expense of fidelity. Suppose thus that $m\ge 1$ additional copies $\ylofi_{n+1}, \dots, \ylofi_{n+m}$ of $\ylofi$ are also observed (without corresponding $\yhifi$'s). In our application of interest (\cref{s:extremes} below), two computer codes of different fidelity are considered for ship motions in random waves, the latter characterized by (random) $x$. For $\yhifi = \yhifi(x)$ and $\ylofi = \ylofi(x)$, we consider, e.g., the maxima of some ship motion over certain time interval given the same waves determined by $x$.

In the same setting, if some quantity of interest (QoI, for short) concerning $\yhifi$ is of interest, one could expect to estimate it better by leveraging the low-fidelity data of $\ylofi$, especially when $\yhifi$ and $\ylofi$ exhibit strong dependence. Better estimation refers to smaller variability (higher efficiency) of the estimator. A well-known and fundamental example of a QoI is the mean of $\yhifi$, that is, $\mu_1 = \bbE \yhifi$. A celebrated multi-fidelity Monte Carlo (MFMC) estimator (e.g., \cite{peherstorfer:2016optimal}) of $\mu_1$ is 
\begin{align}
\label{e:intro-mf-mu}
\widehat\mu_{1,{\rm mf}} = \ybaro + \alpha \left( \ybart - \ybarto \right),
\end{align}
where $\alpha \in \bbR$ and $(\overline{\yjifi})_k = \frac{1}{k}\sum_{i=1}^k \yjifi_i$. The estimator $\widehat \mu_{1,{\rm mf}}$ is unbiased (i.e.\ $\bbE \widehat \mu_{1, {\rm mf}} = \mu_1$). 
An optimal choice of $\alpha$ is to minimize $\var(\widehat\mu_{1,{\rm mf}})$ (see \cref{s:preliminaries} below). 
For this choice of $\alpha = \alpha_{\rm opt}$ and when $\corr(\yhifi, \ylofi)\neq 0$, one always has 
\begin{align}
\label{e:intro-var-bl}
\var(\widehat\mu_{1,{\rm mf}}) < \var(\widehat\mu_{1,{\rm bl}}),
\end{align}
where $\widehat\mu_{1, {\rm bl}} = \ybaro$ is the baseline estimator based just on the high-fidelity data and the efficiency of $\widehat\mu_{1,{\rm mf}}$ over $\widehat\mu_{1,{\rm bl}}$ in terms of \eqref{e:intro-var-bl} increases as the correlation between $\yhifi$ and $\ylofi$ tends to $1$ or $-1$.

\subsection{Goals and contributions of this work}
In the MF setting above, we are interested here in MFMC estimation frameworks when QoIs are estimated through parametric statistical models. For example, in our application of interest, if $\yhifi$ refers to the maximum of some ship motion over large time interval, it is customary from extreme value theory to model its distribution by a parametric generalized extreme value (GEV) distribution (e.g., \cite{coles:2001}).
A QoI in this example could be an exceedance probability for $\yhifi$ of some fixed critical threshold; the probability will depend on the model parameters. How can the exceedance probability be estimated more efficiently having the low-fidelity data on $\ylofi$, with $\ylofi$ and $\yhifi$ being dependent? We note that the critical threshold is typically large enough so that the exceedance probability cannot be estimated directly (as the sample proportion of exceedances) and the standard MFMC method as in \eqref{e:intro-mf-mu} do not apply directly either.
In this work, we propose and study MF estimators supposing that $\yhifi$ and possibly $\ylofi$, or even $(\yhifi, \ylofi)$, are described by parametric statistical models.

Though the motivation above concerns QoIs, we recast the problem as that of MF estimation of parameters of the distribution of $\yhifi$; this then leads naturally back to QoIs. 
We consider several such MF estimators: JML, the joint maximum likelihood estimator, requiring a parametric model for the joint distribution of $\yjoint$; MoM, the moment estimator, requiring a parametric model for $\yhifi$ only and the parameters expressed as moments of functions of $\yhifi$; MML, the marginal maximum likelihood estimator, requiring parametric models separately for the marginal distributions of $\yhifi$ and $\ylofi$.
Note the different parametric assumptions on high- and low- fidelity data.
The JML estimation is expected to be most efficient but asks for a joint parametric model.
The MoM estimation involves a parametric model for $\yhifi$ only but requires parameters expressed as moments, typically at the expense of efficiency. The MML estimation attempts to strike a balance, by requiring the parametric models for $\yhifi$ and $\ylofi$ separately, but its efficiency remains to be understood better and possibly improved. We also note that the JML estimation is certainly not new, and the MoM estimation adapts available MFMC approaches. The MML, on the other hand, seems to be more original.

We examine the asymptotic efficiencies of the MF estimators above and compare them to baselines (when based on the high-fidelity data alone) for several parametric families of distributions: Gaussian, Gumbel (a special case of GEV) and Bernoulli. By considering these specific distributions, we both illustrate and raise important points concerning our methods, including: there are no differences among the three MF estimators in the Gaussian case (with the exception of the MoM estimator of the variance parameter), but substantial differences are observed in the Gumbel case. Implications of our methods on QoIs are somewhat straightforward, but are also potentially of greater relevance in practice. 
Additional topics, including extensions to a general number of sources and how cost considerations can be incorporated into the proposed framework, are also discussed.
We note that, partly for the motivating applications mentioned above, our parametric distributions and QoIs are relevant to extreme value problems, although our approaches are general and accommodate other parametric distributions.

\subsection{Related work}
The MFMC estimator \eqref{e:intro-mf-mu} appears across a range of areas, sometimes under different names, and its exact origins seem difficult to pinpoint. It is also referred to as an approximate control variate (ACV) estimator, where $\ybart$ is used as an approximation of the control variate mean $\bbE \ylofi$. 
In this sense, in a traditional control variates (CV) method (\cite{NELSON1987}), the optimal $\alpha$ is the least-squares coefficient for regressing $\yhifi-\bbE \yhifi$ on $\ylofi-\bbE \ylofi$, leading to an estimator with reduced variance. More recently, such variance reduction techniques have been applied widely in MF settings,
from polynomial chaos-based surrogate models (\cite{Yang2023}) to multi-level constructions arising from numerical discretizations, including the multi-level Monte Carlo (MLMC) and multi-level multi-fidelity (MLMF) estimators (\cite{Giles2015,geraci2017mfml}).
When extended to multiple low-fidelity models, MF estimation involves additional design considerations.
One such consideration is how to allocate samples across fidelity levels under a fixed computational budget (e.g., \cite{peherstorfer:2016optimal}).
Another direction relates to the dependence structures of model evaluations, which may be coupled (e.g., via shared inputs), nested, or independently obtained (e.g., \cite{gorodetsky2020}). See \cref{s:discussion:mvr} for additional details.

In parallel, closely related data structures have been studied in the statistical literature, often framed as semi-supervised learning (SSL) with partially labeled data or missing outcomes.
Here, $\yhifi$ corresponds to the target variable (available only for a subset of data), while $\ylofi$ acts as the predictor (available for all samples). 
Within this framework, recent studies have investigated the asymptotic properties of SSL estimators under different estimation objectives.
\cite{Zhang2016SemisupervisedIG} developed a semi-supervised least squares estimator with a focus on mean estimation, which closely resembles the MFMC estimator \eqref{e:intro-mf-mu}.
\cite{Chakrabortty2018ssl} and \cite{Azriel2022ssl} considered a linear regression setting and the associated parameter estimation problem.
\cite{Chakrabortty2018ssl} used nonparametric imputation to handle missing labels, while \cite{Azriel2022ssl} proposed CV- and ACV-type estimators by recasting regression coefficient estimation as a mean estimation problem to achieve variance reduction.
Moreover, \cite{Song2023ssl} proposed M-estimators for SSL by extending the objective function through a projection-based approach.
A notable difference from the MF formulation is that, in SSL, multiple predictors are observed jointly as a single vector, making multivariate extensions more straightforward; for example, regression-based frameworks often utilize multivariate least-squares regression for estimating the variance-reducing coefficients (\cite{Zhang2016SemisupervisedIG,Chakrabortty2018ssl, Azriel2022ssl}). 

Beyond mean estimation, several extensions of the MF framework have been considered.
One extension focuses on modeling extreme values or rare events.
In the SSL context, for example, \cite{Ahmed2025} addresses inference for tail quantities, leveraging the tail dependence structures to reduce the asymptotic variance of extreme value index and quantile estimators.
Studies in the MF setting have combined the MFMC estimator with importance sampling schemes to construct efficient estimators for failure probabilities (\cite{KRAMER2019}, \cite{pham2022ensemble}).
Another direction closely related to this work concerns MF estimation of functions and multivariate outputs. While our focus is on MF estimation of parametric distribution (density) functions, the related strand of literature considers this task in the non-parametric setting. Multivariate outputs (for example, the distribution function at multiple argument values) are studied in \cite{Dixon2024,Menhorn2024}. Similar ideas are combined with splines or other interpolation operators in estimation of whole functions in \cite{Krumscheid2018}, \cite{AyoulGuilmard2023}.
A key distinction is that our parametric approach estimates the entire 
distribution through a finite-dimensional parameter, enabling 
extrapolation to regions where direct estimation is infeasible and offering greater statistical efficiency when the model is 
correctly specified.

\vspace{1em}
The rest of the paper is organized as follows. \Cref{s:preliminaries} expands on the problem setup. \Cref{s:approach} introduces the three MF parameter estimation methods, and \cref{s:example} examines these estimators across several parametric models. \Cref{s:discussion} includes further discussion related to our results, and \cref{s:extremes} presents the application of our methods to extremal quantities. \Cref{s:conclusions} concludes.
Proofs and some other materials are moved to the supplementary document.

\section{Preliminaries}
\label{s:preliminaries}
As described in \Cref{s:introduction}, 
we consider the following two data settings:
\begin{align}
    \label{e:data:baseline}
    \textit{Baseline}:
    &\quad \yhifi_1, \dots, \yhifi_n, \\
    \label{e:data:mf}
    \textit{MF}: 
    &\quad \big(\yhifi_1, \ylofi_1\big), \dots, \big(\yhifi_n, \ylofi_n\big), ~\ylofi_{n+1}, \dots, \ylofi_{n+m}.
\end{align}
Again, $\yhifi$ represents a `true' or high-fidelity output, and $\ylofi$ a surrogate or low-fidelity version of $\yhifi$. We think of $\yhifi$ and $\ylofi$ as being dependent, for example, resulting from the same underlying randomness conditions. 

The distributions of $\yhifi$ and $\ylofi$ are assumed to be specified in a parametric way. We distinguish between the following two specifications:
\begin{align}
	\textit{ Marginal specification}:  &  ~\bbP (Y^{(j)}\leq y_j) = F^{(j)}_{\btheta_j}(y_j) ; \label{e:prelim-cdf-marg} \\
	\textit{ Joint specification}:  & ~\bbP (\yhifi\leq y_1, \ylofi\leq y_2) = F_\bfeta(y_1, y_2). \label{e:prelim-cdf-joint}
\end{align}
Here, the marginal c.d.f.'s $F_{\btheta_j}^{(j)}$ depend on parameters $\btheta_j\in \bbR ^{d_j}$ and similarly for the joint c.d.f.\ $F_\bfeta$ depending on a vector parameter $\bfeta$, which we further assume to include the marginal parameters $\btheta_1, \btheta_2$ and another parameter $\btheta_{1,2}$ for the dependence of $\yhifi$ and $\ylofi$ as 
$\bfeta = (\btheta_1, \btheta_2, \btheta_{1,2})$.
We shall generally be interested in estimation of the parameter $\tone$ of the  model from the MF data
; see \Cref{s:approach} below. 
Some parametric MFMC methods considered below rely on marginal specifications while others require the joint specification. The marginal specification is naturally more robust to any misspecification in the form of dependence between $\yhifi$ and $\ylofi$. We denote the true parameter values by $\btheta_j^\ast, \tot^\ast, \bfeta^\ast$.

When they exist, the p.d.f.\ and p.m.f.\ are denoted, for $j=1,2$, by $f_{\btheta_j}(y_j) = \partial_{y_j} F_{\btheta_j}^{(j)}(y_j)$, $p_{\btheta_j}(y_j) = F_{\btheta_j}^{(j)}(y_j) - F_{\btheta_j}^{(j)}(y_j-)$, $f_{\bfeta}(y_1, y_2) = \partial_{y_1 y_2}^2 F_{\boldsymbol{\eta}}(y_1, y_2)$, $p_{\bfeta}(y_1, y_2) = \Delta_{y_2}\!\left(\Delta_{y_1} F_{\bfeta}(y_1, y_2)\right)$,
where $\Delta_{y_i}G{(\boldsymbol{y})} := G(\boldsymbol{y}) - G(\boldsymbol{y}_{i-})$, with $\boldsymbol{y}=(y_1, y_2)$, $\boldsymbol{y}_{1-} = (y_{1}-, y_2)$ and $\boldsymbol{y}_{2-} = (y_1, y_{2}-)$.

For reference below, we also consider the case where the vector parameter $\tone \in \bbR ^ {d_1}$ can be expressed in terms of suitable population expectation (moment), that is,
\begin{eqnarray}
\label{e:prelim-moment}
	\mbox{\it Moment formulation} & : &  \tone = \bfg\left(\bbE \bfh(\yhifi)\right),
\end{eqnarray}
where $\bfh = (h_1, \dots, h_{d_1}): \bbR \rightarrow \bbR ^{d_1}$, $\bfg: \bbR ^{d_1} \rightarrow \bbR ^{d_1}$ are known deterministic functions and the relation \eqref{e:prelim-moment} is interpreted elementwise. Moment-based estimation will naturally be used in the case \eqref{e:prelim-moment}. For notational simplicity, we also set: for $l = 1, \dots, d_1$,
\begin{align}
\label{e:prelim-y}
\zhifi_l = h_l(\yhifi),~ \zhifi_{l,i} = h_l(\yhifi_{i}), ~\mu_{Z,l} = \bbE \zhifi_l,
\end{align}
so that ${\bmu}_Z := \bbE \bzhifi = \bbE \bfh(\yhifi)$ reads elementwise $\mu_{Z,l} = \bbE \zhifi_l $ and $\btheta_1 = \bfg(\bmu_Z)$.

\begin{example}\label{ex:mom}
Several parametric models are studied in \cref{s:example}, including the Gaussian distribution \(\yhifi \sim \mathcal{N}(\mu_1,\sigma_1^2)\) with mean $\mu_1$ and variance $\sigma_1^2$. The parameters of this distribution have the moment formulation \eqref{e:prelim-moment} with
\begin{equation}
\begin{split}
\label{e:bvn-mom-form}
\left( \begin{array}{c}
\mu_1 \\ \sigma_1^2
\end{array}
\right)
= \begin{pmatrix}
\bbE \yhifi \\ 
\bbE\big(\yhifi\big)^2 - \big(\bbE \yhifi \big)^2
\end{pmatrix}
&= \left(
\begin{array}{c}
g_1\Big(\bbE h_1(\yhifi), \bbE h_2(\yhifi)\Big)\\
g_2\Big(\bbE h_1(\yhifi), \bbE h_2(\yhifi)\Big)
\end{array}
\right),
\end{split}
\end{equation}
where $h_1(y) = y, h_2(y) = y^2, g_1(u_1, u_2) = u_1$ and $g_2(u_1, u_2) = u_2-u_1^2$.
Furthermore, the notation around \eqref{e:prelim-y} becomes:
$\zhifi_1 = \yhifi$, $\zhifi_2 = (\yhifi)^2$, $\mu_{Z, 1} = \bbE \yhifi$, $\mu_{Z, 2} = \bbE (\yhifi)^2$, so that $\mu_1 = \mu_{Z, 1}$ and $\sigma_1^2 = \mu_{Z, 2} - \mu_{Z, 1}^2$. For the joint specification \eqref{e:prelim-cdf-joint}, we shall assume in \cref{s:example} that $\big(\yhifi, \ylofi\big)$ follows a bivariate Gaussian distribution. In that case, $\tone = (\mu_1, \sigma_1^2)^\top$ and $\ttwo = (\mu_2, \sigma_2^2)^\top$ are the marginal mean and variance parameters, and $\theta_{1,2} = \rho$ is the correlation parameter.
\end{example}

In the example above, we are interested in ways of estimating parameters $\mu_1, \sigma_1^2$ given the MF data, assuming either the marginal or joint specification. As the example has the moment formulation, the MFMC estimator \eqref{e:intro-mf-mu} can naturally be adapted.
But could JML do better assuming the joint specification? This is one question studied in \cref{s:example} for the Gaussian but also other parametric distribution families.

Regarding the MFMC estimator in \eqref{e:intro-mf-mu}, we also note for further reference that 
\begin{align}
\label{e:intro-mf-var}
\var(\widehat\mu_{1,{\rm mf}}) = \frac{1}{n}\var\big(\yhifi\big) + \frac{m}{n(n+m)}\left \{ \alpha^2\var \big(\ylofi\big) - 2 \alpha\cov\big(\yhifi, \ylofi\big) \right\},
\end{align}
which is minimized with the choice of
\begin{align}
\label{e:intro-alpha-opt}
\alpha_{\rm opt} = \frac{\cov\big(\yhifi, \ylofi\big)}{\var\big(\ylofi\big)} = \left( \frac{\var\big(\yhifi\big)}{\var\big(\ylofi\big)}\right)^{1/2}\hspace{-1em} \cdot \corr\big(\yhifi, \ylofi\big),
\end{align}
\begin{align}
\label{e:intro-var-opt}
\left.\var(\widehat\mu_{1,{\rm mf}})\right|_{\alpha = \alpha_{\rm opt}} = \frac{\var\big(\yhifi\big)}{n}\left( 1 - \frac{m}{n+m}\corr\big(\yhifi, \ylofi\big)^2\right).
\end{align}
The last expression explains \eqref{e:intro-var-bl} and the statements made around it.

\section{Parametric MFMC estimation methods}
\label{s:approach}

With the parametric framework described in \cref{s:preliminaries}, we consider several methods to estimate $\tone$, the parameter of $\yhifi$.
We start with several baselines and the joint maximum likelihood, and then present MF alternatives that have computational or other advantages. 
\Cref{tab:mfmc-methods} summarizes the key MF estimation methods discussed in this section that utilize the MF data \eqref{e:data:mf}, along with their definitions and distributional assumptions.
The efficiency gains of JML over the alternatives will be studied for several parametric models in \cref{s:example} below.

\begin{table}[h]
\centering
\caption{Acronyms and key features of parametric MFMC methods}
\label{tab:mfmc-methods}
\resizebox{0.98\textwidth}{!}{\begin{tabular}{lllll}
\toprule
Acronym & Full name & Definition & Specification & MFMC type \eqref{e:intro-mf-mu} \\
\midrule
JML     & Joint Maximum Likelihood  & \eqref{e:mfmc-mle-def}  & joint \eqref{e:prelim-cdf-joint} & No\\
MoM     & Moment Multi-Fidelity & \eqref{e:mom-first}--\eqref{e:mom-second}& marginal \eqref{e:prelim-cdf-marg} for $F^{(1)}_{\tone}$ & Yes \\
MML     & Marginal ML Multi-Fidelity  & \eqref{e:mfmc-mml-first}--\eqref{e:mfmc-mml-second} & marginal \eqref{e:prelim-cdf-marg} & Yes \\
\bottomrule
\end{tabular}}
\end{table}

\subsection{Baselines} 
\label{s:approach:bl}
As a natural baseline in our setting, we consider the maximum likelihood (ML) estimator of $\tone$ based only on the high-fidelity data $\dhifi$ in \eqref{e:data:baseline}.
Assuming the density exists, the estimator is ${\widehat{\btheta}}_{1,{\rm bl, ml}} = \argmax{\tone} {~\prod_{i=1}^n f_{\tone}\big(\yhifi_i\big)} = \argmax{\tone} \sum_{i=1}^n \ell_{\tone}\big(\yhifi\big)$,
where $\ell_{\btheta} = \log f_{\btheta}$.
As we are concerned about the asymptotic variance of the estimators, we expect under regularity assumptions that 
\begin{align}
    \lim_n n\var\big(\htheta_{1,{\rm bl, ml}}\big) =\left. \left( - \bbE \frac{\partial^2 \log f_{\tone}\big(\yhifi\big)}{\partial\tone^2} \right)^{-1} \right|_{\tone = \tstarh} =: \bSigma_{\rm bl, ml} .
    \label{e:mfmc-bl-var}
\end{align}
(When $\tone$ is a vector, \eqref{e:mfmc-bl-var} is understood in the matrix sense, with $\var$ interpreted as a covariance matrix.)
In the moment formulation \eqref{e:prelim-moment}, another natural baseline is the moment estimator
obtained by replacing the population expectations with sample averages.

\begin{example}
\label{ex:base}
For \cref{ex:mom}, the baseline estimators of $\tone = \big(\mu_1, \sigma_1^2\big)^\top$ derived from the ML estimation and the moment formulation coincide, and are given by
\begin{align}
\label{e:ex:bvn:base}
({\htheta}_{1, {\rm bl}})_n = 
     \left( \frac{1}{n} \sum_{i=1}^n \yhifi_i, ~ \frac{1}{n}\sum_{i=1}^n \big(\yhifi_i - \ybaro\big)^2 \right)^\top.
\end{align}
\end{example}

\subsection{Joint maximum likelihood estimation}
\label{s:approach:ml}
At one end of the spectrum in terms of expected efficiency, we have joint maximum likelihood (JML) estimation. 
We use the abbreviation JML rather than ML to distinguish it from the marginal maximum likelihood estimation.
Assuming densities exist, this corresponds to:
\begin{align}
\label{e:mfmc-mle-def}
    (\htheta_{1, {\rm jml}}, \htheta_{2, {\rm jml}}, \htheta_{1,2,{\rm jml}}) &= \argmax{(\tone, \ttwo, \tot)}  \prod_{i=1}^n f_{(\tone, \ttwo, \tot)} \big(\yhifi_i, \ylofi_i\big) \prod_{i=n+1}^{n+m} f_{\ttwo}\big(\ylofi_i\big).
\end{align}

The asymptotic variance of \eqref{e:mfmc-mle-def} will naturally depend on the asymptotic behavior of $n,m \rightarrow \infty$. As we are interested in the situations where there is a large amount of low-fidelity data $\ylofi_i$, the case $m \gg n$ is of particular interest. Simplifying this further, suppose $m$ is effectively infinite so that $\ttwo$ is actually known.
We shall refer to this as:
\begin{eqnarray}
\mbox{Case }m=\infty & : &  \ttwo^\ast \text{ is known.} \label{e:mfmc-case-m} 
\end{eqnarray}
In this case, 
\begin{align}
    (\htheta_{1, {\rm jml}}, \htheta_{1,2, {\rm jml}}) = \argmax{(\tone, \tot)} \prod_{i=1}^n f_{(\tone, \ttwo^\ast, \tot)} \big(\yhifi_i, \ylofi_i\big),
    \label{e:mfmc-mle-joint}
\end{align}
and we expect that, with $\ell_{\btheta} = \log f_{\btheta}$,
\begin{align}
    \lim_n n\var(\htheta_{1,{\rm jml}},\htheta_{1,2,{\rm jml}}) = \left.\left( - \bbE \frac{\partial^2 \ell_{(\tone, \ttwo^\ast, \tot)}\big(\yhifi, \ylofi\big)}{\partial(\tone, \tot)^2} \right)^{-1}\right|_{\tone=\tstarh, \tot = \tot^\ast} =: \bSigma_{\rm jml} .
    \label{e:mfmc-mle-var}
\end{align}
The case $m=\infty$ thus allows comparing the asymptotic variances \eqref{e:mfmc-bl-var} and \eqref{e:mfmc-mle-var} since they both scale as $n^{-1}$.

\subsection{Moment multi-fidelity estimation}
\label{s:approach:mom}

With the moment formulation \eqref{e:prelim-moment}, 
the parameter $\tone$ could be estimated componentwise, employing the standard MFMC approach in \eqref{e:intro-mf-mu}:
\begin{align}
\label{e:mom-first}
(\htheta_{1, {\rm mom}})_k &= g_k \left(\bhbaro + \balpha_k \odot \Big(  \bhbart - \bhbarto \Big) \right)\\
\label{e:mom-second}
&= g_k\left(\balpha_k \odot \bhbart + \Big(  \bhbaro - \balpha_k \odot \bhbarto \Big) \right),
\end{align}
where $\overline{\bfh(X^{(j)})}_n = n^{-1}\sum_{i=1}^{n} \bfh(X_i^{(j)})$, $\balpha_k = (\alpha_{k,1}, \dots, \alpha_{k, d_1})^\top \in \bbR^{d_1}$, and $\odot$ denotes the Hadamard (elementwise) product. The optimal vector $\balpha_k \in \bbR^{d_1}$ is obtained for each \(k=1, \dots, d_1,\) to minimize the variance of \((\htheta_{1, {\rm mom}})_k\). That is, different components of \(\htheta_{1, {\rm mom}}\) may be associated with different optimal coefficients.

With the notation \eqref{e:prelim-y} and defining $\zlofi_l = h_l(\ylofi),~ \zlofi_{l,i} = h_l(\ylofi_{i})$ analogously, \eqref{e:mom-first}--\eqref{e:mom-second} can be interpreted as applying MFMC
elementwise to the moment components $Z_l$, such that $\mu_{Z,l} = \bbE \zhifi_l$ is estimated as
\begin{align}
\label{e:mfmc-mom-element}
\widehat\mu_{Z,l} = \alpha_{k,l} \zbart + \Big(\zbaro - \alpha_{k,l} \zbarto \Big).
\end{align}
We sometimes write $\widehat\bmu_Z(\balpha_k)$ and $\hmu_{Z,l}(\alpha_{k,l})$ to underscore their dependence on $\alpha$.
Similarly to \eqref{e:intro-mf-var}, the covariance structure of $\widehat\bmu_Z$
is given by (with the variance obtained when $r=s$)
\begin{equation}
\begin{split}\label{e:mfmc-moment-Cov-mu}
\cov(\widehat\mu_{Z,r},\widehat\mu_{Z,s}) 
= \frac{\cov(\zhifi_r, \zhifi_s)}{n+m} + \frac{m}{n(n+m)} \cov\Big(\zhifi_r - \alpha_{k,r}\zlofi_r, \zhifi_s - \alpha_{k,s}\zlofi_s \Big).
\end{split}
\end{equation}

To obtain an optimal $\balpha_k$, instead of minimizing $\var(\widehat\mu_{Z,l})$ for each $l$, we suggest to directly minimize the variance of $(\htheta_{1,{\rm mom}})_k = g_k(\widehat\bmu_Z)$ using the delta method.
Specifically, setting 
$\bfG_k := \nabla g_k (\bmu_Z) \in \bbR^{d_1}$ 
for notational simplicity, we have
\begin{align}
\label{e:mfmc-mom-var-tilde}
\var\left((\htheta_{1,{\rm mom}})_k\right) \approx \bfG_k^\top\var(\widehat\bmu_Z) \bfG_k =: \widetilde{\var} \left((\htheta_{1,{\rm mom}})_k\right).
\end{align}
Here,
the $d_1\times d_1$ symmetric matrix $\var(\widehat\bmu_Z)$ has entries specified in 
\eqref{e:mfmc-moment-Cov-mu}, each a quadratic function of 
$\alpha_{k,1},\dots,\alpha_{k,d_1}$.
The optimal $\balpha_k$ is then obtained by solving
\begin{align}
\label{e:mfmc-mom-alpha-l}
\balpha_{k, {\rm opt}} = \argmin{\balpha}\widetilde{\var}\left((\htheta_{1,{\rm mom}})_k\right),
\end{align}
which reduces to solving the system of $d_1$ equations,
$\partial_{\alpha_{k,l}}\widetilde{\var}\big((\htheta_{1,{\rm mom}})_k\big) = 0$, $l=1, \dots, d_1$.
When $\bfG_k$ is estimated independently of $\balpha_k$, this system
is linear. The case $d_1=2$ is considered in Appendix~A of the supplementary material. 
The optimal $\balpha_{k, {\rm opt}}$ can be substituted back into the estimator to obtain $(\htheta_{1,{\rm mom}})_k=g_k(\widehat\bmu_Z(\balpha_{k, {\rm opt}}))$, ensuring that each optimal $\balpha_{k, {\rm opt}}$ is chosen to minimize the variance of each element of $\htheta_{1,{\rm mom}}$.

In the case \eqref{e:mfmc-case-m} 
and with optimal $\balpha$'s from \eqref{e:mfmc-mom-alpha-l}, we expect for the limiting variances that
\begin{align}
\label{e:mfmc-mom-Sig}
\lim_{n \rightarrow \infty} n \widetilde{\var}(\htheta_{1,{\rm mom}}) &= \bSigma_{\rm mom} = (\Sigma_{{\rm mom},l, k})_{l, k = 1, \dots, d_1},
\end{align}
with $\Sigma_{{\rm mom},l, k}$ approximated as
\begin{equation}
\label{e:mfmc-moment-var}
\lim_{n \rightarrow \infty} n \widetilde{\cov}\left((\htheta_{1,{\rm mom}})_l, (\htheta_{1,{\rm mom}})_k\right) = \bfG_l^\top \left(\lim_{n \rightarrow \infty} n\cov(\widehat\bmu_Z(\balpha_{l, {\rm opt}}), \widehat\bmu_Z(\balpha_{k, {\rm opt}}))\right) \bfG_k.
\end{equation}

\begin{example}
\label{ex:mom-mf}
Continuing with \cref{ex:mom}, the MoM estimator for \((\mu_1,\sigma_1^2)\) is:
\begin{align*}
\widehat\mu_{1, {\rm mom}} &= \ybaro + \alpha_{1} \left(\ybart - \ybarto\right) ,\\
\widehat\sigma^2_{1, {\rm mom}} &=g_2\left(
\zbaroone + \alpha_{2,1} \Big(\zbartone - \zbartoone\Big),
\zbarotwo + \alpha_{2,2} \Big(\zbarttwo - \zbartotwo
\Big)
\right),
\end{align*}
where $Z_1^{(j)} = h_1(\yjifi) = \yjifi$, $Z_2^{(j)}=h_2(\yjifi)= (\yjifi)^2$ and $g_2(u_1, u_2) = u_2 - u_1^2$.
The optimal choice of $\alpha_1$ and $\balpha_2 = (\alpha_{2,1}, \alpha_{2,2})^\top$ will be given in \cref{s:ex-bvn}.
\end{example}

\subsection{Marginal maximum likelihood multi-fidelity estimation}
\label{s:approach:mml}
The moment estimator \eqref{e:mom-first}--\eqref{e:mom-second} is naturally expected to be generally less efficient than the JML estimator \eqref{e:mfmc-mle-def}; see illustrations in \cref{s:example}. On the other hand, the moment estimator does not need the joint specification \eqref{e:prelim-cdf-joint}; it is based on the marginal specification \eqref{e:prelim-cdf-marg} for $j=1$ only. A natural question in that regard is whether a multi-fidelity estimator could be considered that would assume only the marginal specification \eqref{e:prelim-cdf-marg} but where the model parameters be estimated through maximum likelihood rather than moments. We consider here one such possibility.

Assume that the low- and high-fidelity parametric models are the same, that is, $\btheta_j = \btheta \in \bbR^{d_1}$ and $F_{\btheta_j}^{(j)} (y)= F_{\btheta} (y)$. 
The true parameters $\tstarh, \tstarl$ do not need to be the same.
Let $(\htheta_{j,{\rm ml}})_n$ be the ML estimator of $\btheta_j^\ast$ based on the data $\yjifi_1, \dots, \yjifi_n$, and similarly for $(\htheta_{2, {\rm ml}})_{n+m}$ based on the data $\ylofi_{1}, \dots, \ylofi_{n+m}$. For the desired multi-fidelity estimator, consider
 \begin{align}
 \label{e:mfmc-mml-first}
 \htheta_{1,{\rm mml}} &= \hto + \bbeta \odot \big( \htt -  \htto \big)\\
 \label{e:mfmc-mml-second}
 &= \bbeta \odot \htt  + \big( \hto - \bbeta \odot \htto\big),
 \end{align}
where $\bbeta = (\beta_1, \dots, \beta_{d_1})^\top \in \bbR^{d_1}$.
We shall refer to \eqref{e:mfmc-mml-first}--\eqref{e:mfmc-mml-second} as the marginal maximum likelihood (MML) multi-fidelity estimator.

The idea behind \eqref{e:mfmc-mml-first}--\eqref{e:mfmc-mml-second} is as follows.
Let $\htheta_n$ denote the maximum likelihood estimator that solves the likelihood equation $\sum_{i=1}^n \dot{\ell}_\btheta(Y_i) = 0$,
where $\dl_\btheta (y) :=  \partial_\btheta\log f_\btheta (y) \in \bbR^{d_1}$ and $f_\btheta(y)$ is a parametric p.d.f.. Set also $\ddl_\btheta(y) := \partial^2_{\btheta} \log f_\btheta (y) \in \bbR^{d_1\times d_1}$.
Under regularity conditions, we expect $\sqrt{n}(\htheta_n - \btheta^\ast)$ to be asymptotically normal with mean zero and covariance matrix $\left.(\bbE \Ddot{\ell}_\btheta)^{-1}\bbE_\btheta \dot{\ell}_\btheta\dot{\ell}_\btheta^\top(\bbE \Ddot{\ell}_\btheta^\top)^{-1}\right|_{\btheta=\btheta^\ast} = (\bfI_{\btheta^\ast})^{-1}$,
with the Fisher information matrix defined as
$\bfI_\btheta = \bbE\dl_\btheta\dl_\btheta^\top = -\bbE \ddl_\btheta.$
It can also be shown (e.g.\ \cite{Vaart_1998}) that
\begin{align}
\label{e:mfmc-mle-asymp}
\sqrt{n}(\htheta_n - \btheta^\ast) = \bfI_{\btheta^\ast}^{-1}\frac{1}{\sqrt{n}}\sum_{i=1}^n \dl_{\btheta^\ast}(Y_i) + o_p(1).   
\end{align}
In our MF setting, the result \eqref{e:mfmc-mle-asymp} implies
$(\htheta_{j,{\rm ml}})_n - \btheta_{j}^\ast \approx n^{-1} \sum_{i=1}^n \bfI_{\btheta^\ast_j}^{-1}\dl_{\btheta^\ast_j}(\yjifi_i)$,
and hence \eqref{e:mfmc-mml-first} can be thought of as
\begin{align}
\htheta_{1,{\rm mml}} &= \tstarh +  (\hto - \tstarh) + \bbeta \odot \big( (\htt - \tstarl) - (\htto - \tstarl) \big) \notag\\
&\approx \tstarh + \frac{1}{n}\sum_{i=1}^n \bIinvh \dl_{\tstarh}(\yhifi_i)  + \bbeta \odot \left( \frac{1}{n+m} \sum_{i=1}^{n+m} \bIinvl \dl_{\tstarl}(\ylofi_i) - \frac{1}{n} \sum_{i=1}^n \bIinvl \dl_{\tstarl}(\ylofi_i) \right) \notag\\
\label{e:mfmc-mml-approx}
&= \tstarh + \bhobaro + \bbeta \odot \Big( \bhtbart - \bhtbarto \Big),
\end{align}
where 
\begin{align}
\label{e:mfmc-mml-h}
\bhhifi(y) = \bfI_{\btheta_1^\ast}^{-1} \dl_{\btheta_1^\ast}(y), ~\bhlofi(y) = \bfI_{\btheta_2^\ast}^{-1} \dl_{\btheta_2^\ast}(y).
\end{align}

The derived form \eqref{e:mfmc-mml-approx} leads to several insights regarding the estimator $\htheta_{1,{\rm mml}}$. Note that \eqref{e:mfmc-mml-approx} resembles the form \eqref{e:mom-first} of the MoM estimator. Note also that $\bbE \bfh^{(1)}(\yhifi) = \bbE \bfh^{(2)}(\ylofi) = \boldsymbol{0}$, so that the MoM estimator is written for zero means. Its construction is also quite special. Unlike \eqref{e:mom-first}, the functions $\bhhifi, \bhlofi$ depend on the true parameters $\tstarh, \tstarl$. The term $\bhobaro$ has the special property that its variance is the smallest since it results from the ML estimation. The variance can furthermore be reduced through the multi-fidelity component $\bhtbart - \bhtbarto$. In that regard, we consider the choice of the ML estimator $\htheta_{2,{\rm ml}}$ as a control variate and the resulting function $\bhlofi(y)$ natural, at least for the following reason. In the `extreme' case when $\yhifi$ and $\ylofi$ are perfectly dependent, taking $\bbeta=\boldsymbol{1}$ in \eqref{e:mfmc-mml-first}--\eqref{e:mfmc-mml-second} or \eqref{e:mfmc-mml-approx}, leads to $\htheta_{1,{\rm mml}} = (\htheta_{2,{\rm ml}})_{n+m}$, which corresponds to the most efficient estimation of $\tone$ in this scenario.
On the other hand, when $\yhifi$ and $\ylofi$ are independent, no efficiency gain is expected with MF estimation and the choice $\bbeta = \boldsymbol{0}$ in \eqref{e:mfmc-mml-first}--\eqref{e:mfmc-mml-second} or \eqref{e:mfmc-mml-approx} leads to the most efficient estimator $\htheta_{1,{\rm mml}} = (\htheta_{1, {\rm ml}})_{n}$ in this scenario.

The argument \eqref{e:mfmc-mml-approx} also suggests an optimal $\bbeta$ to use. Indeed, $\beta_{l,{\rm opt}}$ can be derived elementwise by minimizing $\var(\widehat\mu_{Z,l})$ in \eqref{e:mfmc-moment-Cov-mu} with $\zhifi_l = h_{l}^{(1)}(\yhifi)$ and $\zlofi_l = h_{l}^{(2)}(\ylofi)$.
As in \eqref{e:intro-alpha-opt}--\eqref{e:intro-var-opt}, this leads to 
\begin{align}
\label{e:mfmc-mml-beta}
\beta_{l,{\rm opt}} &= \argmin{\beta_l} \var(\hmu_{Z,l}) = 
\frac{\cov(\zhifi_l, \zlofi_l)}{\var(\zlofi_l)}, \\
\label{e:mfmc-mml-var-mn}
\left.\var(\widehat\mu_{Z,l})\right|_{\beta_l = \beta_{l, {\rm opt}}} &= \frac{\var(\zhifi_l)}{n}\left(1-\frac{n}{n+m} \corr\left(\zhifi_l, \zlofi_l\right)^2\right).
\end{align}
Under the case \eqref{e:mfmc-case-m}, we expect that
\begin{align}
\label{e:mfmc-mml-var}
\lim_{n \rightarrow \infty} n \var(\htheta_{1,{\rm mml}}) &= \bSigma_{\rm mml} = (\Sigma_{l, k})_{l, k = 1, \dots, d_1}
\end{align}
holds with $\Sigma_{l, k}$ defined as
\begin{align}
\label{e:mfmc:mml-var-mu}
n\var(\widehat\mu_{Z,l}) &= \var(\zhifi_l) \left( 1 - \text{Corr}\left(\zhifi_l, \zlofi_l\right)^2 \right) =: \Sigma_{l,l},\\
\label{e:mfmc:mml-cov-mu}
n\cov(\widehat\mu_{Z,l}, \widehat\mu_{Z,k}) &= \cov( \zhifi_{l} - \beta_{l,{\rm opt}} \zlofi_{l} , \zhifi_{k} - \beta_{k,{\rm opt}} \zlofi_{k} ) =: \Sigma_{l, k}.
\end{align}

\begin{remark}
\label{rmk:mfmc-mml}
Note that the expected limiting variance in \eqref{e:mfmc:mml-var-mu}, expressed as
\begin{align*}
\var\big(\hhifi_l(\yhifi)\big) \left(1-\corr\big(\hhifi_l(\yhifi), \hlofi_l(\ylofi)\big)^2\right)=: T_1 (1-T_2),
\end{align*}
has two components $T_1$ and $T_2$. 
The term $T_1$ results from the ML estimation for $\yhifi$ and will be the smallest one could achieve since it results from the ML estimation as noted following (\ref{e:mfmc-mml-h}).
On the other hand, the question remains open whether $T_2$ could be made larger, that is, whether there is another choice replacing $\htheta_{2,{\rm ml}}$ in \eqref{e:mfmc-mml-first}--\eqref{e:mfmc-mml-second}, and expressed only through the marginal information,
with the resulting function $\bhlofi$ so that $|\corr\big(\hhifi_l(\yhifi), \hlofi_l(\ylofi)\big) |$ is as large as possible. As motivated above, the current approach \eqref{e:mfmc-mml-first}--\eqref{e:mfmc-mml-second} has appealing features but it is generally not optimal (e.g.\ see \cref{s:ex-bvgumbel} below).
\end{remark}

\section{Examples and numerical illustrations}
\label{s:example}

This section presents illustrative examples to evaluate the efficiency of the proposed MF estimators.
We consider several distributional settings where the joint structure is fully specified. 
\Cref{s:ex-bvn} analytically demonstrates the efficiency of MF estimators for a bivariate Gaussian distribution.  
\Cref{s:ex-bvgumbel,s:ex-ber} examine the performance of the estimators under bivariate Gumbel and Bernoulli distributions, respectively, highlighting how it varies across dependence levels. 
In the numerical illustrations, it is assumed that MF estimators utilize optimal coefficients ($\balpha$ or $\bbeta$) to maximize their efficiency.
The reproducible R code used in this study is included in the supplementary material.

\subsection{Bivariate Gaussian distribution}
\label{s:ex-bvn}
An instructive example is the case when the distribution of $\yhifi$ and $\ylofi$ is assumed to be Gaussian with the location and scale parameters $\mu_j$ and $\sigma_j$, $j=1,2$, and the correlation coefficient $\rho$.
The parameters are $\btheta_j = (\mu_j, \sigma_j^2)^\top$ and $\bfeta = (\btheta_1, \btheta_2, \rho)^\top$. 
See \cref{ex:base,ex:mom-mf} for the moment formulation and baseline estimators for the Gaussian distribution.

The various estimation methods discussed in \cref{s:approach} can be applied to estimate $\tone$ using MF approaches.
In this context, \cref{prop:bvn:joint-mle} considers the JML estimation as defined in \eqref{e:mfmc-mle-def}, which sets a benchmark by providing an optimal estimator with the integration of additional low-fidelity data. 
A proof based on solving the likelihood equation is given in Appendix~B of the supplementary material, 
while Appendix~E provides an alternative proof highlighting the connection to the MFMC estimator.

\begin{proposition}
\label{prop:bvn:joint-mle}
Assume $(\yhifi, \ylofi) $ follows the bivariate Gaussian distribution. The JML estimator defined in \eqref{e:mfmc-mle-def} is given by 
\begin{align}
\label{e:bvn-m1}
\widehat\mu_{1,{\rm jml}} &= \ybaro +  \widehat \alpha\left(\ybart-\ybarto\right),\\
\label{e:bvn-m2}
\widehat\mu_{2,{\rm jml}} &= \ybart, \\
\label{e:bvn-s1}
\widehat \sigma_{1,{\rm jml}}^2
&= \frac{1}{n}\sum_{i=1}^{n}(\yhifi_i- \ybaro)^2 + \widehat \alpha^2\left(\widehat \sigma_{2,{\rm jml}}^2 - \frac{1}{n}\sum_{i=1}^{n}(\ylofi_i-\ybarto)^2  \right ),\\
\label{e:bvn-s2}
\widehat \sigma_{2,{\rm jml}}^2 &= \frac{1}{n+m}\sum_{i=1}^{m+n}(\ylofi_i-\ybart)^2,\\
\label{e:bvn-rho}
\widehat\rho_{\rm jml} &= \frac{\widehat\sigma_{2,{\rm jml}}}{\widehat\sigma_{1,{\rm jml}}} ~\widehat \alpha,\qquad \widehat\alpha = \frac{\sum_{i=1}^n (\yhifi_i-\ybaro)(\ylofi_i-\ybarto)}{\sum_{i=1}^n(\ylofi_i-\ybarto)^2}.
\end{align}
\end{proposition}

The MoM estimator of $(\mu_1, \sigma_1^2)$ was considered in \cref{ex:mom-mf}, leaving the presentation of the optimal coefficients $\alpha_1$ and $\balpha_2 = (\alpha_{2,1}, \alpha_{2,2})^\top$ to this section. They are given by
\begin{align}
\label{e:bvn-m1-mom-alpha}
\alpha_{1, {\rm opt}} &= \rho^\ast\frac{ \sigma_1^\ast}{\sigma_2^\ast}, \quad
\balpha_{2, {\rm opt}} = \left(  \frac{ \mu^\ast_2}{\mu^\ast_1}\Big(\rho^\ast\frac{ \sigma_1^\ast}{\sigma_2^\ast}\Big)^2,
\Big(\rho^\ast\frac{ \sigma_1^\ast}{\sigma_2^\ast}\Big)^2
\right)^\top,
\end{align}
as derived in Appendix~A of the supplementary material. 
The MoM estimation \eqref{e:mom-first} becomes
\begin{align}
\label{e:bvn-mom-m1}
\widehat \mu_{1,{\rm mom}} &= \ybaro + \Big(\frac{\rho^\ast \sigma^\ast_1}{\sigma^\ast_2}\Big) \left( \ybart - \ybarto \right), \\
\label{e:bvn-mom-s1}
\widehat\sigma_{1,{\rm mom}}^2 &= \big(\overline{(\yhifi)^2}\big)_n +  \Big(\frac{\rho^\ast \sigma^\ast_1}{\sigma^\ast_2}\Big)^2 \left( \big(\overline{(\ylofi)^2}\big)_{n+m} - \big(\overline{(\ylofi)^2}\big)_n \right)\notag \\
& \quad - \left(\ybaro + \frac{\mu_2^\ast}{\mu_1^\ast}\Big(\frac{\rho^\ast \sigma^\ast_1}{\sigma^\ast_2}\Big)^2 \left( \ybart - \ybarto\right)\right)^2.
\end{align}
When $\rho^\ast\frac{ \sigma_1^\ast}{\sigma_2^\ast}$ is estimated as $\widehat\alpha$ in \eqref{e:bvn-rho}, the MoM estimator \eqref{e:bvn-mom-m1} coincides with the JML estimator \eqref{e:bvn-m1}, whereas \eqref{e:bvn-mom-s1} differs from \eqref{e:bvn-s1}.

The MML estimator from \cref{s:approach:mml} involves 
\begin{equation*}
\begin{aligned}
\dl_{\btheta_1^\ast}(y) &= \begin{pmatrix}
    \frac{y-\mu_1^\ast}{(\sigma_1^\ast)^2} \\
    \frac{(y-\mu_1^\ast)^2-(\sigma_1^\ast)^2}{2(\sigma_1^\ast)^4}\end{pmatrix} , ~
\bfI_{\btheta_1^\ast}  =  \begin{pmatrix}
    \frac{1}{(\sigma_1^\ast)^2} & 0 \\
    0 & \frac{1}{2 (\sigma_1^\ast)^4}
\end{pmatrix},~
\bhhifi(y) = \begin{pmatrix}
    y-\mu_1^\ast \\
    {(y-\mu_1^\ast)^2}-{(\sigma_1^\ast)^2}\end{pmatrix}.
\end{aligned}
\end{equation*}
Based on this $\bfh^{(1)}(y)$, the optimal $\bbeta = (\beta_1, \beta_2)^\top$ is determined by \eqref{e:mfmc-mml-beta} as
\begin{align}
\beta_{1,{\rm opt}} &= \frac{\cov(\yhifi-\mu^\ast_1, \ylofi-\mu^\ast_2)}{\var (\ylofi-\mu^\ast_2)} = \rho^\ast\frac{ \sigma^\ast_1}{\sigma^\ast_2},\\
\beta_{2,{\rm opt}} &= \frac{\cov((\yhifi-\mu^\ast_1)^2, (\ylofi-\mu^\ast_2)^2)}{\var\left((\ylofi-\mu^\ast_2)^2\right)} = \left(\rho^\ast \frac{\sigma^\ast_1}{\sigma^\ast_2}\right)^2.
\end{align}
Then, the MML estimation \eqref{e:mfmc-mml-first}--\eqref{e:mfmc-mml-second} becomes
\begin{align}
\label{e:bvn-mml-m1}
\widehat \mu_{1,{\rm mml}} &= ({\widehat \mu}_{1, {\rm bl}})_n + \rho^\ast \frac{\sigma^\ast_1}{\sigma^\ast_2} \left( ({\widehat \mu}_{2, {\rm bl}})_{n+m} - ({\widehat \mu}_{2, {\rm bl}})_n \right), \\
\label{e:bvn-mml-s1}
\widehat\sigma_{1,{\rm mml}}^2 &= ({\widehat \sigma^2}_{1, {\rm bl}})_n + \left(\rho^\ast \frac{\sigma^\ast_1}{\sigma^\ast_2}\right)^2 \left(({\widehat \sigma^2}_{2, {\rm bl}})_{n+m} - ({\widehat \sigma^2}_{2, {\rm bl}})_n \right ), 
\end{align}
where $({\widehat \mu}_{j, {\rm bl}})_{k} = (\overline{\yjifi})_k$ and $({\widehat \sigma^2}_{j, {\rm bl}})_{k} = k^{-1}\sum_{i=1}^k \big(\yjifi_i - (\overline{\yjifi})_k\big)^2$ for $j=1,2$ denote the baseline ML estimators based on a sample of size $k$. With $\rho^\ast\frac{\sigma_1^\ast}{\sigma_2^\ast}$ estimated as $\widehat\alpha$ in \eqref{e:bvn-rho}, \eqref{e:bvn-mml-m1} and \eqref{e:bvn-mml-s1} align with \eqref{e:bvn-m1} and \eqref{e:bvn-s1}.

\subsection{Bivariate Gumbel distribution}
\label{s:ex-bvgumbel}
In contrast to \cref{s:ex-bvn}, we consider here a distribution where more substantial differences among the MF estimators exist.
Take a bivariate Gumbel distribution with c.d.f.\ 
\begin{equation}\label{e:blockmax-G}
    F_\bfeta(y_1,y_2) = \exp \left\{-\left(z_1 + z_2 \right)A\left(\frac{z_1}{z_1 + z_2} \right) \right\},\quad z_j = \exp\left\{-\frac{y_j - \mu_j}{\sigma_j}\right\}, ~ j=1,2,
\end{equation}
with location and scale parameters $\mu_j\in\bbR$ and $\sigma_j>0$, respectively, and certain function $A:[0,1]\rightarrow [0,1]$. Here, $\btheta_j = (\mu_j, \sigma_j)^\top$, and $\bfeta$ represents the vector of parameters $(\btheta_1, \btheta_2)$ along with parameters for the function $A$, if any.

The function $A$ is called the Pickands dependence function and has the following properties: $A: [0, 1] \mapsto [0, 1]$ is convex, $\max(t, 1-t) \le A(t) \le 1$, and $A(0) = A(1) = 1$. Under \eqref{e:blockmax-G}, the marginal c.d.f.'s of $\yhifi$ and $\ylofi$ are given by the univariate Gumbel distributions $F_{\btheta_j}^{(j)}(y_j) = \exp \left\{- z_j \right\}$, $j=1,2$.
The function $A$ determines the dependence structure between $\yhifi$ and $\ylofi$, with $A \equiv 1$ representing independence. The case $A(t) = \max(t, 1-t)$ leads to perfect positive dependence.
In practice, $A$ in \eqref{e:blockmax-G} takes parametric form. For example, the logistic Pickands function is given by
\begin{equation}\label{e:blockmax-G-log}
    A(t) = (t^{1/r} + (1-t)^{1/r})^r,
\end{equation}
for some parameter $r \in (0, 1]$. Here, $r$ is a one-parameter gauge of dependence, interpolating between independence ($r=1$) and perfect positive dependence ($r\to0$).
Bivariate Gumbel distributions arise as possible limits of normalized bivariate maxima of blocks of bivariate observations. See \cite{balakrishnan2009}, Chapter 12, for more information.

In this context, we explore various estimators discussed in \cref{s:approach} and compare their asymptotic variances as a measure of their efficiency.
It is known that the maximum likelihood estimation generally performs better than the moment estimation for the Gumbel distribution. 
We consider here the case where both $\mu_1$ and $\sigma_1$ are unknown; 
the known $\sigma_1$ case is treated in Appendix~D of the supplementary material.
The moment formulation for $\tone$ is written as (see \cite{johnson1995continuous}, Chapter 22)
\begin{equation}
\begin{split}
\label{e:bvg-mom-form}
\tone 
= \bfg\left(\bbE \zhifi_1, \bbE \zhifi_2\right),
\quad \bfg(z_1, z_2)  = \begin{pmatrix}
z_1 - \gamma \sqrt{\frac{6}{\pi^2}(z_2 -  z_1^2)} \\
 \sqrt{\frac{6}{\pi^2}(z_2 -  z_1^2)}
\end{pmatrix},
\end{split}
\end{equation}
where $Z_1^{(j)} = h_1(\yjifi) = \yjifi,~ Z_2^{(j)}=h_2(\yjifi)= (\yjifi)^2$ and
\begin{align}
\label{e:bvg-mom-dgdy}
\bfG:= 
\left(
\begin{array}{c}
\nabla g_1(\bmu_Z)^\top\\
\nabla g_2(\bmu_Z)^\top
\end{array}\right)
= \begin{pmatrix}
    1 + \gamma \sqrt{\frac{6}{\pi^2}} \frac{\mu_{Z,1}}{\sqrt{\mu_{Z,2} -  \mu_{Z,1}^2}} & -\gamma \sqrt{\frac{6}{\pi^2}} \frac{1}{2\sqrt{\mu_{Z,2} -  \mu_{Z,1}^2}}  \\
    -\sqrt{\frac{6}{\pi^2}} \frac{\mu_{Z,1}}{\sqrt{\mu_{Z,2} -  \mu_{Z,1}^2}} & \sqrt{\frac{6}{\pi^2}} \frac{1}{2\sqrt{\mu_{Z,2} -  \mu_{Z,1}^2}}
\end{pmatrix}.
\end{align}
The baseline moment estimator is obtained as
\begin{align}
\label{e:bvg-bl-mom-sig}
&\htheta_{1,{\rm bl,mom}} = \bfg\Big(\zbaroone, \zbarotwo\Big), \quad
n\widetilde{\var}(\widehat\btheta_{1,{\rm bl,mom}}) = \bfG \bSigma \bfG^\top ,
\end{align}
where $\bSigma$ is the $2\times 2$ covariance matrix with entries $(\bSigma)_{ij} = \cov(\zhifi_i, \zhifi_j)$ for $i,j=1,2.$
The MoM estimator is given as
\begin{align}
\label{e:bvg-mom-sig}
(\htheta_{1,{\rm mom}})_k = g_k(\widehat\bmu_Z),\quad \balpha_{k, {\rm opt}} = \argmin{\balpha} ~ \bfG_k^\top \var(\widehat\bmu_Z) \bfG_k, ~k=1,2.
\end{align}
The optimal coefficients could be obtained using Proposition A.1 of the supplementary material. The asymptotic variances are then computed as in \eqref{e:mfmc-moment-var}.
The ML and MML estimators are constructed using (\cite{johnson1995continuous}, Chapter 22)
$$
\dl_{\btheta_j^\ast}(y) = \begin{pmatrix}
\frac{1}{\sigma_j^\ast}\big(1 -\exp{(-z_j)}\big)\\
-\frac{1}{\sigma_j^\ast}\big(1-z_j +z_j \exp{(-z_j)}\big)
\end{pmatrix}, ~ \bfI_{\btheta_j^\ast} = \frac{1}{(\sigma_j^\ast)^2}\begin{pmatrix}
1 & \gamma -1\\
\gamma -1 & (\gamma -1)^2 + \frac{\pi^2}{6}
\end{pmatrix},$$
where $z_j = {(y-\mu_j^\ast)}/{\sigma_j^\ast}$. We then have $\bfh^{(j)}(y) = \bfI_{\btheta_j^\ast}^{-1}\dl_{\btheta_j^\ast}(y)$.

\begin{figure}[t]
\centerline{
\includegraphics[width = .75\textwidth]{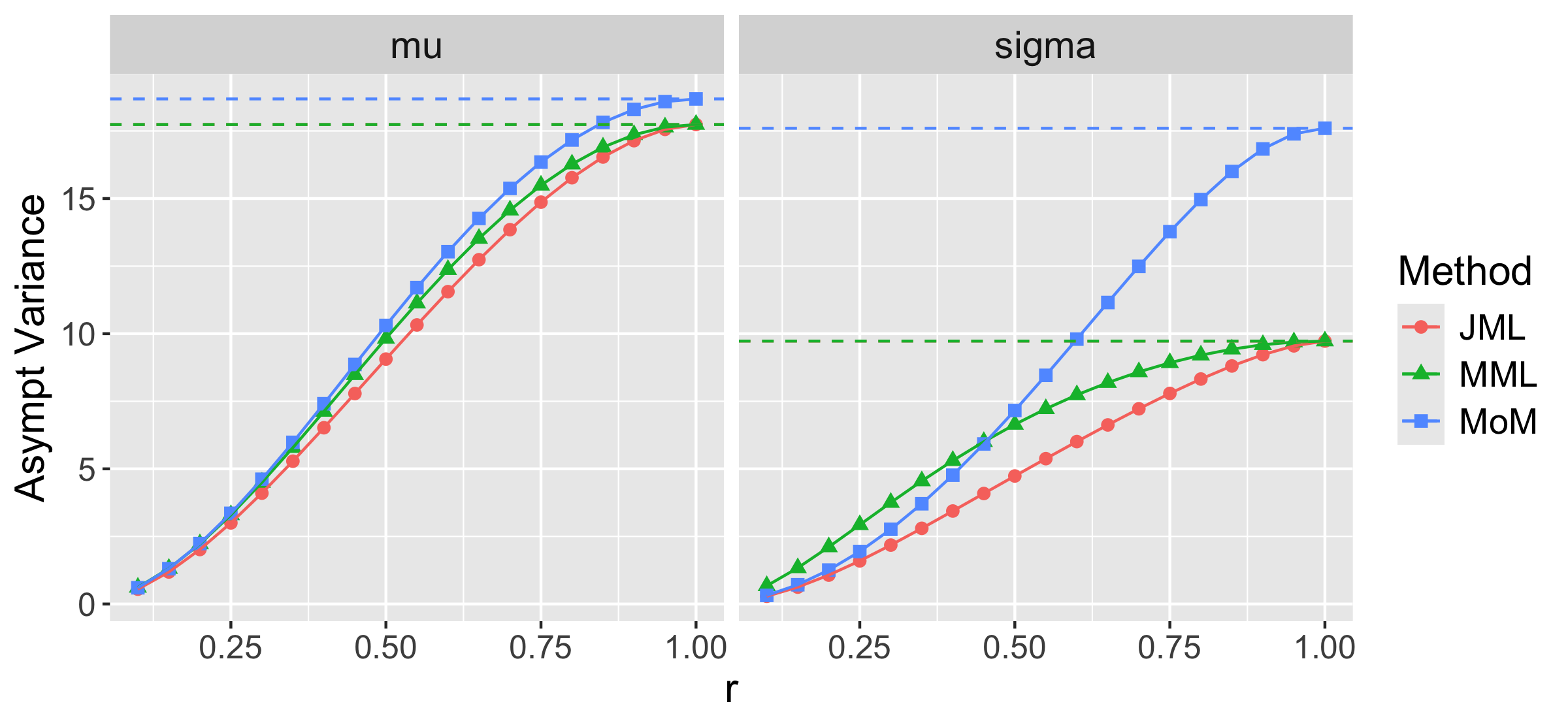}}
\caption{Asymptotic variances of JML (red, circle), MML (green, triangle), MoM (blue, square), and baseline (dashed) estimators for $\mu_1$ and $\sigma_1$ across dependence parameter values $r$ for the bivariate Gumbel distribution.}
\label{f:bvg-unknownS}
\end{figure}
\Cref{f:bvg-unknownS} shows the (rescaled) asymptotic variances of the three MF estimators that incorporate low-fidelity outputs across varying dependence levels $r$. The true parameters are set as $\tstarh = (2,4)^\top,~ \tstarl = (2,1)^\top,$ and $r$ ranges from $0.1$ to $1$ in increment of $0.05$.
Dashed lines indicate the baseline variances, which depend only on the marginal high-fidelity data and are therefore constant in~$r$. 
For $r=1$, representing independence between variables, both the MML and MoM estimators show no variance reduction compared to the respective baselines, indicating no additional information gain from low-fidelity data. 
For~$\mu_1$ (left panel), all three MF estimators show relatively similar performances, with MF estimators under marginal specifications closely matching the JML estimator, whereas substantial differences emerge for $\sigma_1$ (right panel). 
The JML estimator consistently achieves the lowest variance.
As dependence strengthens (smaller~$r$), all MF estimators improve, but at different rates: the MML estimator closely tracks the JML, while the MoM estimator initially performs worse than MML but eventually surpasses it at stronger dependence levels, coming much closer to the variance of the JML estimator. As noted in \cref{rmk:mfmc-mml}, the MML estimator is thus not optimal when only marginal parametric specification is assumed.

\subsection{Bivariate distribution of binary outcomes}
\label{s:ex-ber}
For discrete distributions, an interesting application involves modeling the joint distribution of binary outcomes. We consider two Bernoulli random variables \(\yjifi\):
\begin{align}
\label{e:ex:ber}
\yjifi = \bbone{\{U_j \le p_j\}}, \quad j=1,2,
\end{align}
where \(U_j\) are uniformly distributed over \([0,1]\). This model is relevant, for example, when analyzing exceedance probabilities $p_j = \bbP(W^{(j)} >w_j)$ for some random variables $W^{(j)}$ and thresholds $w_j$.

To model the dependence between $\yhifi$ and $\ylofi$, we use a copula function \(C: [0,1]^2 \to [0,1]\) which links the joint distribution to the marginals via
$F_\bfeta(y_1, y_2) = C\big(F^{(1)}_{\btheta_1}(y_1), F^{(2)}_{\btheta_2}(y_2)\big)$
(\cite{Nelsen2006}). 
In this section, we model the dependence between the latent uniform variables \(U_1\) and \(U_2\) in \eqref{e:ex:ber} using a parametric copula,
$\bbP(U_1 \le u_1, U_2 \le u_2) = C(u_1, u_2; \bphi)$,
where $\bphi \in \Phi$ is a dependence parameter.
We denote $\bfeta = (p_1, p_2, \bphi)^\top$ and define
\begin{equation} \label{e:gamma-def}
    \gamma := \bbP(\yhifi=1, \ylofi=1) = \bbP(U_1 \le p_1, U_2 \le p_2) = C(p_1, p_2; \bphi),
\end{equation}
under which the joint p.m.f. can be written as
\begin{align}
\label{e:ex-ber-pmf}
p_\bfeta(y_1,y_2)
&= \gamma^{y_1y_2} \big(p_1 - \gamma\big)^{y_1(1-y_2)}\big(p_2 - \gamma\big)^{y_2(1-y_1)}  \big(1-p_1 -p_2 + \gamma\big)^{(1-y_1)(1-y_2)}.
\end{align}
Since the likelihood \eqref{e:ex-ber-pmf} depends on the copula parameter $\bphi$ only through \(\gamma\),
we reparametrize to \(\tilde\bfeta=(p_{1},p_{2},\gamma)^\top\).
The attainable range of \(\gamma\) is constrained by the Fréchet–Hoeffding bounds (\cite{Nelsen2006}),
$
  \Gamma(p_{1},p_{2})
  =\{C(p_{1},p_{2};\bphi):\bphi\in\Phi\}
  \subseteq[\max\{0,p_{1}+p_{2}-1\},\min\{p_{1},p_{2}\}].
$
We maximize the log-likelihood over the feasible set $\tilde\Theta
  =\bigl\{(p_{1},p_{2},\gamma):
          0<p_{j}<1,\; \gamma\in\Gamma(p_{1},p_{2})\bigr\}$.
The resulting JML estimators are given in the next proposition;
see Appendix~C of the supplementary material for a proof.
\begin{proposition}
\label{prop:ber:jml}
Assume $\yjifi$, $j= 1,2$, are given by \eqref{e:ex:ber}, whose dependence is governed by a parametric copula $C(u_1,u_2;\bphi)$, $\bphi \in \Phi$. Define the re-parameterization
\(\tilde\bfeta=(p_{1},p_{2},\gamma)^\top\) with $\gamma = C(p_1, p_2;\bphi)$.
Then, the JML estimators of $\tilde \bfeta$ are
\begin{align}
\label{e:ber-p1-jml}
\widehat p_{1,{\rm jml}} &= \ybaro + \frac{\frac{1}{n}\sum_{i=1}^n \yhifi_i\ylofi_i - \ybarto\ybaro}{\ybarto(1-\ybarto)}\left(\ybart-\ybarto
\right), \\
\label{e:ber-p2-jml}
\widehat p_{2,{\rm jml}} &= \frac{1}{n+m} \sum_{i=1}^{n+m} \ylofi_i,
\quad \quad \widehat\gamma_{\rm jml} = \frac{\ybart}{\ybarto}\frac{1}{n}\sum_{i=1}^n \yhifi_i\ylofi_i.
\end{align}
\end{proposition}

On the other hand, the MML and MoM estimators for $p_1$ coincide, leading to the same MFMC estimator.
More specifically, including the baseline estimator, we have
\begin{align}
\label{e:ex-ber-base}
\widehat p_{1, {\rm bl}} &= \frac{1}{n}\sum_{i=1}^n \yhifi_i, ~n\var(\widehat p_{1, bl}) = p_1^\ast(1-p_1^\ast), \\
\label{e:ex-ber-mf}
\widehat p_{1, {\rm mom}} &= \widehat p_{1, {\rm mml}} = \ybaro + \alpha_{\rm opt} \left( \ybart - \ybarto \right),\\
\label{e:ex-ber-mf-var}
n\var(\widehat p_{1, {\rm mom}}) &= n\var(\widehat p_{1, {\rm mml}}) = p_1^\ast(1-p_1^\ast) \left(1-\corr(\yhifi, \ylofi)^2\right).
\end{align}
Utilizing the log-likelihood function $\ell_{p_j}(y) = y\log p_j + (1-y) \log(1-p_j)$, we have $$\dot\ell_{p_j^\ast}(y) =\frac{y}{p_j^\ast} - \frac{1-y}{1-p_j^\ast}, \quad \ddot\ell_{p_j^\ast}(y) = - \frac{y}{(p_j^\ast)^2} - \frac{1-y}{(1-p_j^\ast)^2}, \quad I_{p_j^\ast}^{-1} = p_j^\ast(1-p_j^\ast).$$ Here, the function in \eqref{e:mfmc-mml-h} reduces to a simple structure, $h^{(j)}(y)=y-p_j^\ast$, from which it follows that
$\alpha_{\rm opt} = \beta_{\rm opt} = {\cov(\yhifi, \ylofi)}/{\var(\ylofi)}$. 
With $\alpha_{\rm opt}$ and $\beta_{\rm opt}$ estimated from data, the resulting MML and MoM estimators in \eqref{e:ex-ber-mf} further coincide with the JML estimator in \eqref{e:ber-p1-jml}.
This provides an example where the MF approach under marginal specification achieves the same efficiency as the JML estimator.

\begin{figure}[t]
\centering
\begin{subfigure}{.49\linewidth}
    \includegraphics[width=1\linewidth]{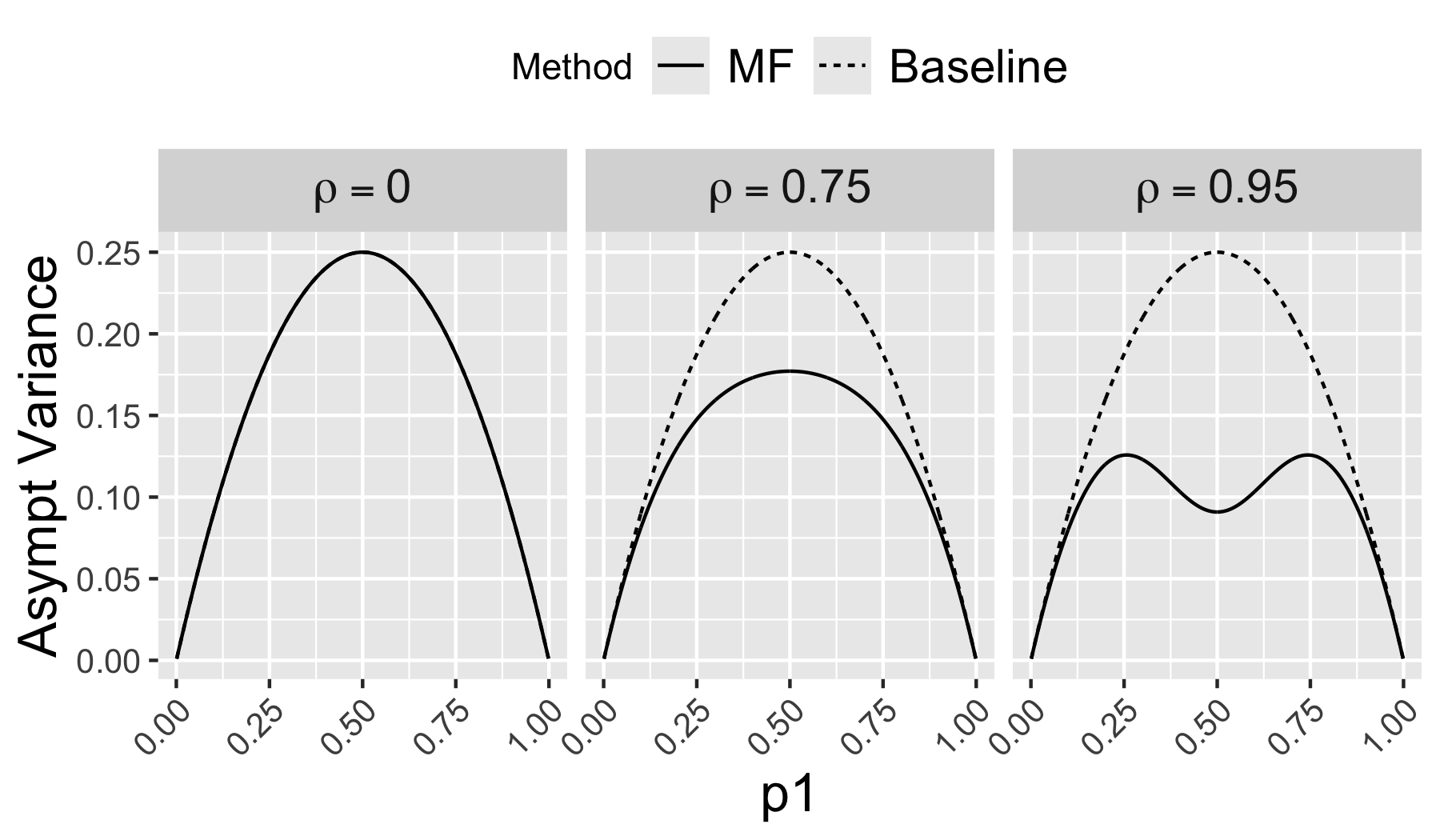}
     \vspace{-3em}
    \caption{Gaussian copula}
    \label{f:ns:f1m2}
\end{subfigure}
\begin{subfigure}{.49\linewidth}
    \includegraphics[width=1\linewidth]{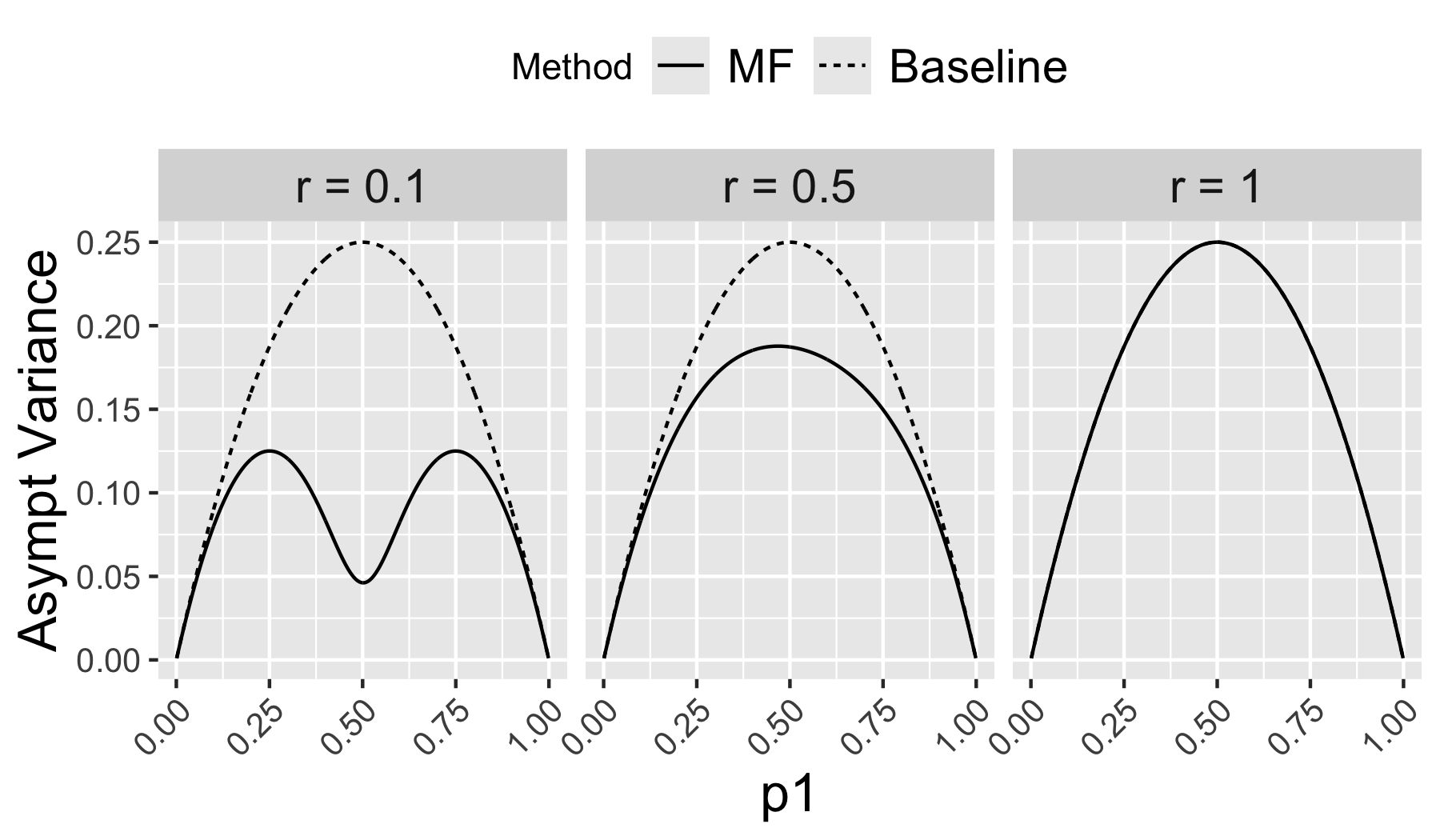}
    \vspace{-3em}
    \caption{Gumbel–Hougaard copula}
    \label{f:ns:f1m1}
\end{subfigure}

\caption{Asymptotic variances of the MF (solid) and baseline (dashed) estimators for $p_1$ for Bernoulli random variables across different dependence levels and copula functions.}
\label{f:ber-variance}
\end{figure}

\Cref{f:ber-variance} compares the asymptotic variances of the baseline and MF estimators for $p_1$ in \eqref{e:ex-ber-base} and \eqref{e:ex-ber-mf-var}, respectively, for two copula functions: the Gaussian copula with each column panel representing a correlation parameter $\rho \in \{0, 0.75, 0.95\}$, as used in \cref{s:ex-bvn}, and the Gumbel-Hougaard copula with a dependence parameter $r \in \{0.1, 0.5, 1\}$, utilized in \cref{s:ex-bvgumbel}.
Each panel compares the asymptotic variances for $p_1$, assuming $p_2^\ast = 0.5$.
When the two sources are independent (i.e., $\rho=0$ or $r=1$), the MF estimator yields no variance reduction relative to the baseline. As dependence strengthens (i.e., larger $\rho$ or smaller $r$), the variance of the MF estimator decreases relative to the baseline, enabling more efficient estimation of $p_1$.

\section{Further discussion and results}
\label{s:discussion}

\subsection{Parametric quantities of interest}
\label{s:discussion:qoi}

The parametric MF methods presented above naturally have implications on how to work with quantities of interest (QoIs, for short) depending on the underlying parametric distributions.
Denote such general QoI as 
\begin{align}
\label{e:qoi}
    q := q(\tone) := \Tilde{q}(F_{\tone}^{(1)}),
\end{align}
where $q$ and $\Tilde{q}$ are used to indicate that the QoI can be viewed as depending on $\tone$ and $\Fhifi_{\tone}$.

\begin{example}
\label{ex:qoi}
In the context of extreme values, consider the Gumbel distribution, that is, $\Fhifi_{\tone}(y_1) = \exp\left\{-\exp\left\{-{(y_1-\mu_1)}/{\sigma_1}\right\}\right\}$, $y_1\in \bbR$,
where $\tone = (\mu_1, \sigma_1)^\top$.
An example of a QoI is the log of exceedance probability
\begin{equation}
\begin{split}
\label{e:qoi-Gumbel-exceedP}
q = \log_{10} \bbP(\yhifi > a_1) &= \log_{10}(1-\Fhifi_{\tone}(a_1)) \\
&= \log_{10}\Big(1- \exp\left\{-\exp\left\{-\frac{a_1-\mu_1}{\sigma_1}\right\}\right\}\Big) =: q(\mu_1, \sigma_1),
\end{split}
\end{equation}
where the target threshold $a_1$ is fixed. Another QoI could be an extreme quantile
\begin{align}
\label{e:qoi-Gumbel-quantile}
q = \inf\left\{y: \Fhifi_{\tone} (y) \ge p_1\right\} = \mu_1 - \sigma_1 \ln(-\ln p_1) =: q(\mu_1, \sigma_1),
\end{align}
for $p_1$ close to $1$ and fixed.
\end{example}

In the traditional setting, corresponding here to having only the high-fidelity data $\yhifi_1, \dots, \yhifi_n$, a natural estimator for the QoI in \eqref{e:qoi} is
\begin{align}
\label{e:qoi:bl}
\widehat q_{\rm bl} = q\big(\widehat \btheta_{1,{\rm bl}}\big),
\end{align}
where $\widehat \btheta_{1,{\rm bl}}$ denotes a baseline estimator (see \cref{s:approach:bl}). When $\widehat \btheta_{1,{\rm bl}}$ is obtained from ML, the estimator \eqref{e:qoi:bl} could be viewed as being ML (\cite{casella2002}, Theorem 7.2.10). In view of these observations, in the MF setting, for the MF estimator of \eqref{e:qoi}, we suggest  
\begin{align}
\label{e:qoi-mf}
\widehat q_{\rm mf} = q\big(\widehat \btheta_{1,{\rm mf}}\big),
\end{align}
where $\widehat \btheta_{1,{\rm mf}}$ is any of the MF estimators of $\tone$, considered in \cref{s:approach}: the JML estimator \eqref{e:mfmc-mle-def}, the MoM estimator \eqref{e:mom-first} or the MML estimator \eqref{e:mfmc-mml-first}.

In the case \eqref{e:mfmc-case-m}, by the delta method and assuming smoothness of $q(\tone)$, we expect for the asymptotic variance that 
\begin{align}
\label{e:qoi-avar}
\lim_{n\rightarrow \infty} n \var(\widehat q_{\rm mf}) = \left. \frac{\partial q}{\partial \tone}^\top \right |_{\tone = \tstarh} \bSigma_{\rm mf} \left. \frac{\partial q}{\partial \tone} \right |_{\tone = \tstarh},
\end{align}
where $\bSigma_{\rm mf}$ is one of the asymptotic covariances in \eqref{e:mfmc-mle-var}, \eqref{e:mfmc-mom-Sig} or \eqref{e:mfmc-mml-var}.
An analogous relation is expected for the baseline estimator $\widehat q_{\rm bl}$ with the limiting covariance matrix in \eqref{e:mfmc-bl-var}. We also note that for the MoM and MML estimation, the coefficients $\balpha$ in (\ref{e:mom-first}) and $\bbeta$ in (\ref{e:mfmc-mml-first}) are chosen to minimize the diagonal entries of $\bSigma_{\rm mf}$ in (\ref{e:qoi-avar}). Another possibility is to choose them to minimize (\ref{e:qoi-avar}) itself, but this is not the approach pursued here. 

\begin{example}
\label{ex:qoi-gumbel}
We continue with \cref{ex:qoi} concerning the Gumbel distribution. \Cref{f:bvg-unknownS} discussed in \cref{s:ex-bvgumbel} presented the asymptotic variances for the considered MF and baseline estimators as functions of the dependence parameter $r$ for fixed parameter values $\tstarh=(\mu_1^\ast, \sigma_1^\ast)^\top = (2,4)^\top$ and $\tstarl=(\mu_2^\ast, \sigma_2^\ast)^\top = (2,1)^\top$. For the estimators of the QoIs \eqref{e:qoi-Gumbel-exceedP} and \eqref{e:qoi-Gumbel-quantile}, their asymptotic variances are now calculated using \eqref{e:qoi-avar}. For \eqref{e:qoi-Gumbel-exceedP}, we have
\[
\left.\frac{\partial q}{\partial \tone}^\top \right |_{\tone = \tstarh} = \frac{\exp\{-z_1-\exp\{-z_1\}\}}{\ln{10}\left(1-\exp\{-\exp\{-z_1\}\}\right)} \left({\frac{1}{\sigma_1^\ast}}, 
{\frac{z_1}{\sigma_1^\ast}} \right), \quad
z_1= \frac{a_1-\mu_1^\ast}{\sigma^\ast_1},
\] and for \eqref{e:qoi-Gumbel-quantile}, \[
\left.\frac{\partial q}{\partial \tone}^\top \right |_{\tone = \tstarh} = \left( 1, -\ln(-\ln p_1) \right).
\] 
\end{example}

\subsection{Multiple sources of information and computational costs}
\label{s:discussion:mvr}
Our focus thus far has been on developing MF estimators leveraging a single low-fidelity source to estimate parameters of the high-fidelity output distribution. Two important topics to consider in our MF setting are: (a) extending these estimators to incorporate multiple sources of information, and (b) addressing the resource allocation problem concerning computational costs. 
For both of these extensions, it is helpful to distinguish between the JML estimator and the MoM and MML estimators, where the latter are based on the standard MFMC formulation given in \eqref{e:intro-mf-mu}. Extensions of MFMC-type estimators to address (a) and (b) have been well studied in the literature and can be adapted to our setting.
In this section, we follow \cite{gorodetsky2020}, and to keep the paper self-contained, we present some of their results.
The extension of the JML estimator requires separate consideration due to its reliance on the joint parametric assumption.

Consider first the extension of MF estimators to multiple sources. The formulation of such extensions depends on the specific structure of the data obtained. For example, consider a general MF scenario with one high-fidelity output $\yhifi$ and $K$ distinct low-fidelity outputs $\ylofi, \dots, \ylofik$. Assume the following dataset is available:
\begin{equation}
\begin{split}
\label{e:discussion-mv-data}
\textit{ Full observations}&: (\yhifi_i, \ylofi_i, \dots, \ylofik_i), ~i = 1, \dots, s_1 ;\\
\textit{Partial observations}&:  (Y_j^{(k)}, \dots, \ylofik_j), ~j = s_{k-1}+1, \dots, s_{k}, ~k= 2, \dots, K+1,
\end{split}
\end{equation}
where $s_{k} = s_{k-1} + m_{k}$ and $s_0 = 0,~ m_1 = n$. MF datasets with other structures could also be considered when $K\ge 2$; see \cite{gorodetsky2020}.

Given this dataset, extending the JML estimator to handle multiple information sources is straightforward, though it explicitly requires assumptions on the joint parametric distribution across sources. The JML estimator can be defined as:
\begin{align}
\hat \bfeta_{\rm jml}
=
\argmax{\bfeta}
\Biggl[
\prod_{i=1}^n
f_{\bfeta}\bigl(\yhifi_i,\dots,\ylofik_i\bigr)\times
\prod_{k=2}^{K+1}\prod_{j=s_{k-1}+1}^{s_k}
f_{\bfeta_k}\bigl(Y_j^{(k)}, \dots, \ylofik_j\bigr)
\Biggr],
\end{align}
where $\bfeta$ denotes the set of parameters for the joint p.d.f. of $(\yhifi, \dots, \ylofik)$ and $\bfeta_k$ those for $(Y^{(k)}, \dots, \ylofik)$. It can be analogously adapted to other data collection structures.

On the other hand, for the MoM and MML estimators which adapt \eqref{e:intro-mf-mu}, it suffices to consider the extensions of \eqref{e:intro-mf-mu} to incorporate multiple information sources.
For example, \cite{gorodetsky2020} noted that the optimal performance can differ between estimators even when using the same dataset \eqref{e:discussion-mv-data}, as in the following cases:
\begin{align}
\label{e:mv:nested}
\widehat \mu_{1, \text{mfmc}} &= \ybaro + \sum_{k=2}^{K+1} \alpha_k \left(
(\overline{Y^{(k)}})_{s_{k}} - (\overline{Y^{(k)}})_{s_{k-1}}
\right),\\
\label{e:mv:acv-mf}
\widehat \mu_{1, \text{acv-mf}} &= \ybaro + \sum_{k=2}^{K+1} \alpha_k \left(
(\overline{Y^{(k)}})_{s_{k}} - (\overline{Y^{(k)}})_{n}
\right),
\end{align}
called the MFMC and ACV-MF estimators, respectively.
The optimal coefficients and corresponding variance for \eqref{e:mv:nested} are given as $\alpha_{k, {\rm opt}} = {\cov(\yhifi, Y^{(k)})}/{\var(Y^{(k)})}$ and 
\begin{align}
\label{e:mvr-nested-var}
\var(\widehat \mu_{1, \text{mfmc}}(\balpha_{\rm opt})) = \frac{\var(\yhifi)}{n}\left(1 - \sum_{k=2}^{K+1} \corr(\yhifi, Y^{(k)})^2 \left( 
\frac{1}{r_{k-1}} - \frac{1}{r_k}\right)\right),
\end{align}
where $r_k = s_k/n$.
The optimal coefficients and associated variance for \eqref{e:mv:acv-mf} are given as 
\begin{align}
\label{e:mvr-acv-var}
\balpha_{\rm opt} = \tilde \bfV^{-1} \bfv,\quad \var(\widehat \mu_{1, \text{acv-mf}}(\balpha_{\rm opt})) = \frac{\var(\yhifi)}{n}\left(1 - \tilde \bfv^\top \tilde \bfV^{-1} \tilde \bfv \right),
\end{align}
where $\tilde \bfv = \bfv / {\sqrt{\var(\yhifi)}}$,
$\bfv = (v_k)_{k=2, \dots, K+1}$, and $\tilde \bfV = (\tilde V_{kl})_{k,l=2, \dots, K+1}$ is symmetric with entries
$v_{k} = {\cov(\yhifi, Y^{(k)})}\left(1 - {1}/{r_k}\right)$ and 
$\tilde V_{kl} = \cov(Y^{(k)}, Y^{(l)})\left(1 - {1}/{r_k}\right)$ for $k \le l$.

This comparison demonstrates that the performance of multivariate MF estimators can vary depending on how they are constructed. For example, for the estimator \eqref{e:mv:nested}, assume without loss of generality that $|\rho_2| \ge |\rho_k|$ for all $2 \le k \le K+1$, where $\rho_k := \corr(\yhifi, Y^{(k)})$. Then, 
we have:
\begin{align*}
\label{e:mvr-nested-var-compare}
n \var(\widehat \mu_{1, \text{mfmc}}(\balpha_{\rm opt})) \ge {\var(\yhifi)}\left(1 - \rho_2^2\right),
\end{align*}
which implies that the estimator cannot outperform the best single low-fidelity case. On the other hand, for the estimator in \eqref{e:mv:acv-mf}, we have:
as $r_k \rightarrow \infty$ for all $2 \le k \le K+1$,
\[
\lim_{n\rightarrow \infty} n\var(\widehat \mu_{1, \text{acv-mf}}(\balpha_{\rm opt})) = \var(\yhifi)(1-\bar{\bfv}^\top \bfV^{-1} \bar{\bfv}),
\]
where 
$\bar{v}_{k} = {\cov(\yhifi, Y^{(k)})}/{\sqrt{\var(\yhifi)}}$ and $V_{kl} = \cov(Y^{(k)}, Y^{(l)})$ for all $2\le k, l \le K+1$. According to \cite{gorodetsky2020}, Section 3, the estimator \eqref{e:mv:acv-mf} can achieve a smaller asymptotic variance than the estimator \eqref{e:mv:nested}, presenting a promising extension of the MoM and MML estimators to the multiple sources.

Turning to the sample allocation problem under a limited computational budget, consider the following general optimization problem:
\begin{align}
\label{e:cost:opt}
\min_{\bfs}  J(\bfs) \quad \text{ subject to } \sum_{k=1}^{K+1} s_k c_k \le C, \quad  s_k \ge s_{k-1} \ge 0
\end{align}
where $c_k$ denotes the per-sample computational cost of the $k$-th fidelity source $Y^{(k)}$, $C>0$ is the total computational budget, and $\bfs=(s_1, \dots, s_{K+1})$ is the vector of sample sizes for each fidelity level. 
The objective function $J(\bfs)$ represents the estimator variance, which depends on the allocation vector $\bfs$. 

In practice, we propose solving this problem for a particular QoI, considered in \cref{s:discussion:qoi}. As in \eqref{e:qoi-avar}, we express the variance through the delta method to take
\begin{align}
\label{e:mvr-alloc-delta}
J(\bfs) = \left. \frac{\partial q}{\partial \tone}^\top \right |_{\tone = \widehat \btheta_1} \var(\widehat \btheta_{1, {\rm mf}}) \left. \frac{\partial q}{\partial \tone} \right |_{\tone = \widehat \btheta_1},
\end{align}
where $\widehat \btheta_{1, {\rm mf}}$ is one of the MF estimators of $\tone$ in \cref{s:approach} and (e.g.\ baseline) $\widehat \btheta_1$ estimates $\btheta_1$. For the MoM estimator, the variance $\var(\widehat \btheta_{1, {\rm mf}})$ can be approximated by plugging \eqref{e:mfmc-mom-alpha-l} into \eqref{e:mfmc-mom-var-tilde} and using \eqref{e:mfmc-moment-Cov-mu} for $\var(\widehat \bmu_Z)$. For the MML estimator, the variance is approximately $\var(\widehat \bmu_Z)$ with its elements in \eqref{e:mfmc:mml-var-mu} and \eqref{e:mfmc:mml-cov-mu}. For the JML estimator, to compute the (finite sample) variance, we revisit the definition \eqref{e:mfmc-mle-def}. The JML estimator $\widehat \bfeta_{\rm jml} :=(\htheta_{1, {\rm jml}}, \htheta_{2, {\rm jml}}, \htheta_{1,2,{\rm jml}})^\top$ is obtained as the solution to the likelihood equations:
\[
S(\bfeta) := \sum_{i=1}^n \nabla_{\bfeta} \ell_{\bfeta}(\yhifi_i, \ylofi_i)
+
\sum_{i=n+1}^{n+m} \nabla_{\bfeta}\ell_{\ttwo}(\ylofi_i)
= \boldsymbol{0},
\]
where $\bfeta = (\tone, \ttwo, \tot)^\top$ and $\nabla_{\bfeta}\ell_{\ttwo}(\ylofi_i) = \left(\boldsymbol{0}^\top, \nabla_{\btheta_2} \ell_{\btheta_2}( \ylofi_i)^\top, \boldsymbol{0}^\top\right)^\top$. We apply a first-order Taylor expansion of $S(\bfeta)$ around the true parameter $\bfeta^\ast$ to get:
\[
\widehat \bfeta_{\rm jml} - \bfeta^\ast \approx - \nabla_{\bfeta} S(\bfeta^\ast)^{-1}S(\bfeta^\ast) \text{ and } \var(\widehat \bfeta_{\rm jml})  \approx \nabla_{\bfeta} S(\bfeta^\ast)^{-1} \var(S(\bfeta^\ast)) (\nabla_{\bfeta} S(\bfeta^\ast)^{-1})^{\top}.
\]

The finite-sample variance for $K \ge 2$ can be derived analogously, following the earlier discussion in this section, and is therefore omitted for brevity. When $J(\bfs)$ represents the variance of a MF estimator targeting $\bbE \yhifi$, the optimization problem for the ACV-MF estimator in \eqref{e:mv:acv-mf} is solved using a local gradient-based method, specifically the interior-point method SLSQP, in \cite{gorodetsky2020}.
We note that a similar approach can also be employed in our setting.

\section{Application to extremes}
\label{s:extremes}
\begin{figure}
\centerline{
\includegraphics[width=2in,height=1.8in]{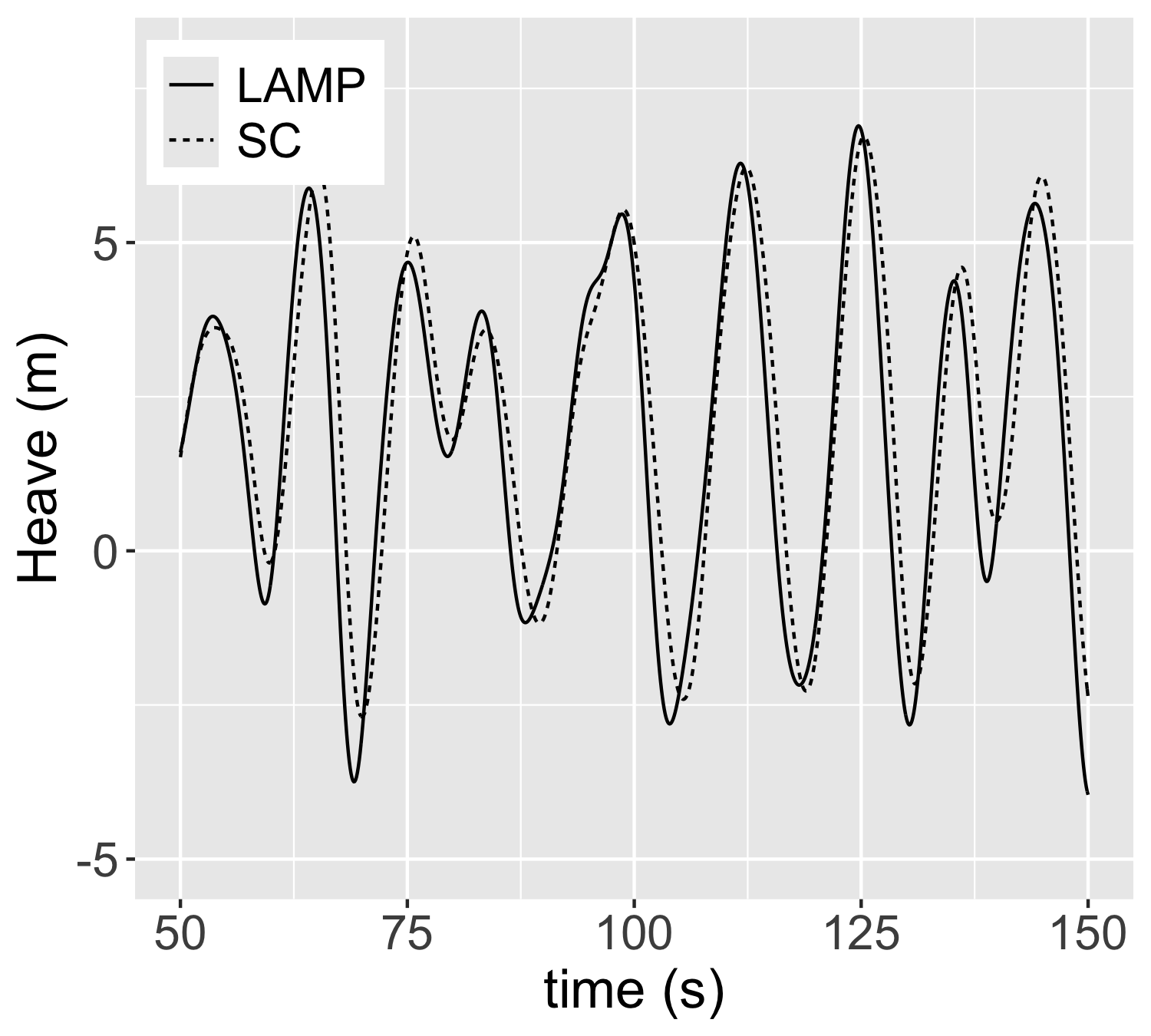}
\includegraphics[width=2.1in,height=1.8in]{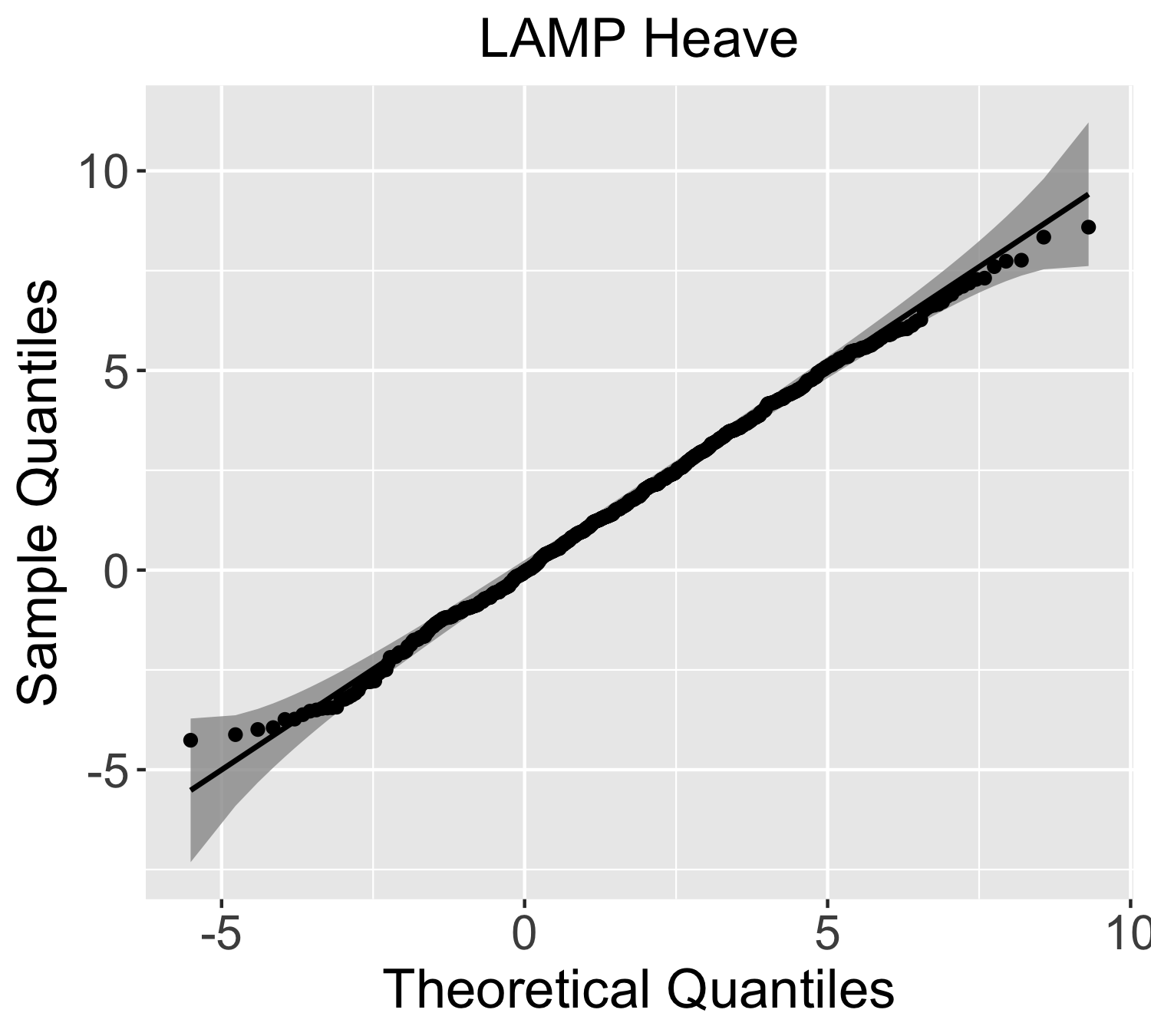}
\includegraphics[width=2.1in,height=1.8in]{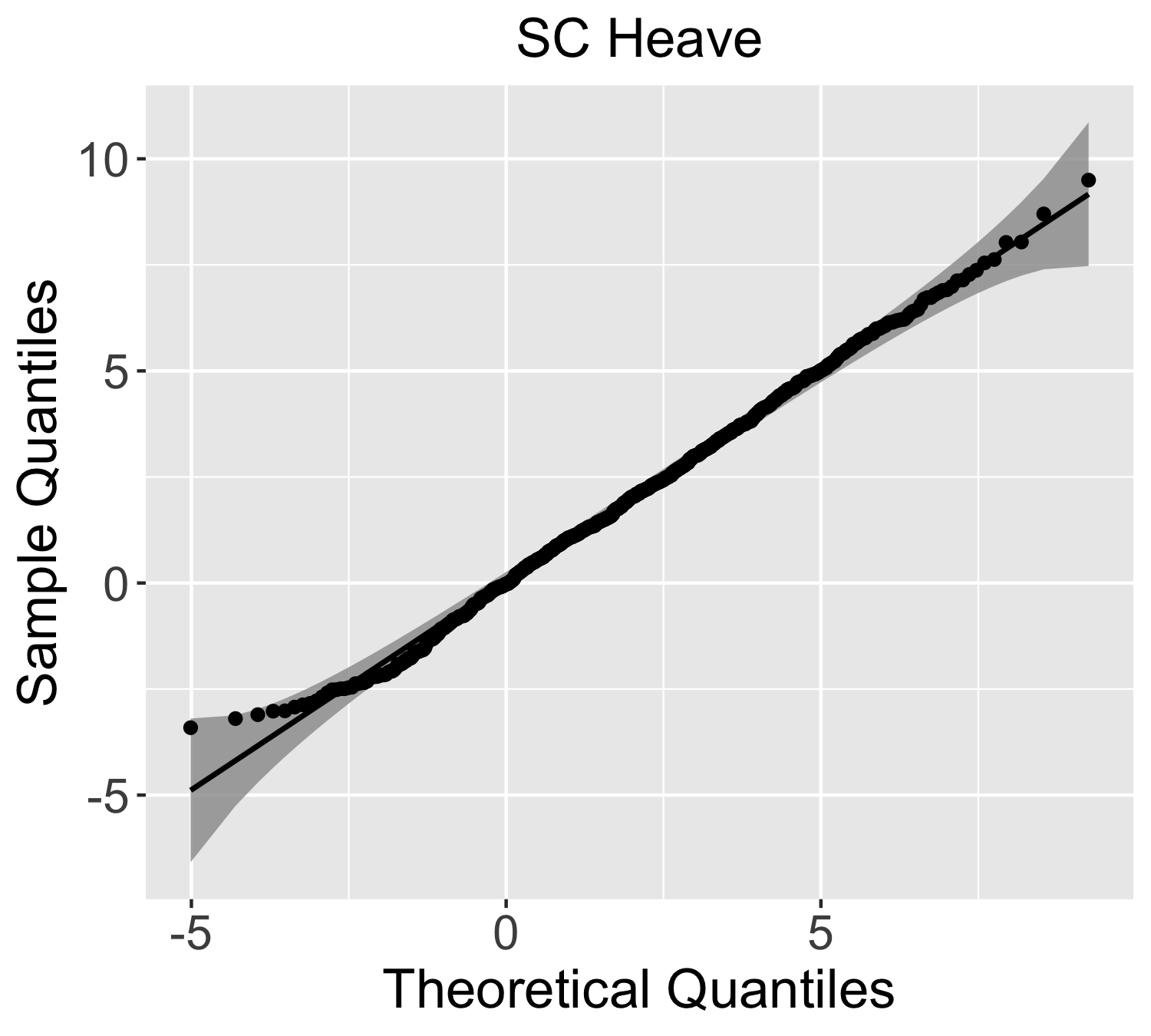}
}
\caption{Left: Heave motion over time from LAMP and SC. Middle/Right: Normal quantile-quantile (QQ) plot for heave motion from LAMP/SC.}	
\label{f:qqplots}
\end{figure}

The application to be used for illustration here concerns modeling ship motions in random waves. Two computer codes widely used by researchers in Naval Architecture are SimpleCode (SC; \cite{weems:wundrow:2013}) and Large Amplitude Motion Program (LAMP; \cite{lin:yue:1991}). The two codes differ in the underlying physics, with LAMP being higher fidelity and SC computationally more efficient. For instance, generating a 30-minute record of ship motions takes about 2-3 seconds with SC, whereas the same process with LAMP can take 15-20 minutes or longer, depending on what outputs are sought. The left plot of \Cref{f:qqplots} shows the heave, one of the motions, of a particular ship in certain conditions generated by the two codes for the same (random) wave excitation over a time window of 100 secs. (Both codes have a number of parameters and options that need to be set, which is not our focus.)

In our MF setting, we focus on the heave motion and generate 30-minute records, considering the record maxima for the two codes as the random variables of interest.
Simulations for each record are associated with a specific random seed $x$ determining the (random) wave excitation for that record. 
The high-fidelity output, $\yhifi = \yhifi(x)$, is the LAMP record heave maximum, and the low-fidelity output, $\ylofi = \ylofi(x)$, is the SC record heave maximum, with the shared underlying random wave excitation determined by the same seed $x$. 
To fit parametric models, a sensible first step is to examine the empirical distributions of data. The middle and right plots of \Cref{f:qqplots} present the normal quantile-quantile (QQ) plots for LAMP and SC heave motions, respectively, each obtained from one 30-minute record (observing heave every 0.1 seconds, resulting in 18,000 data points). These plots suggest that the heave motions follow closely Gaussian distributions, and we expect their record (block) maxima to asymptotically converge to Gumbel distributions.

\begin{figure}[t]
\centerline{
\includegraphics[width=2in,height=1.8in]{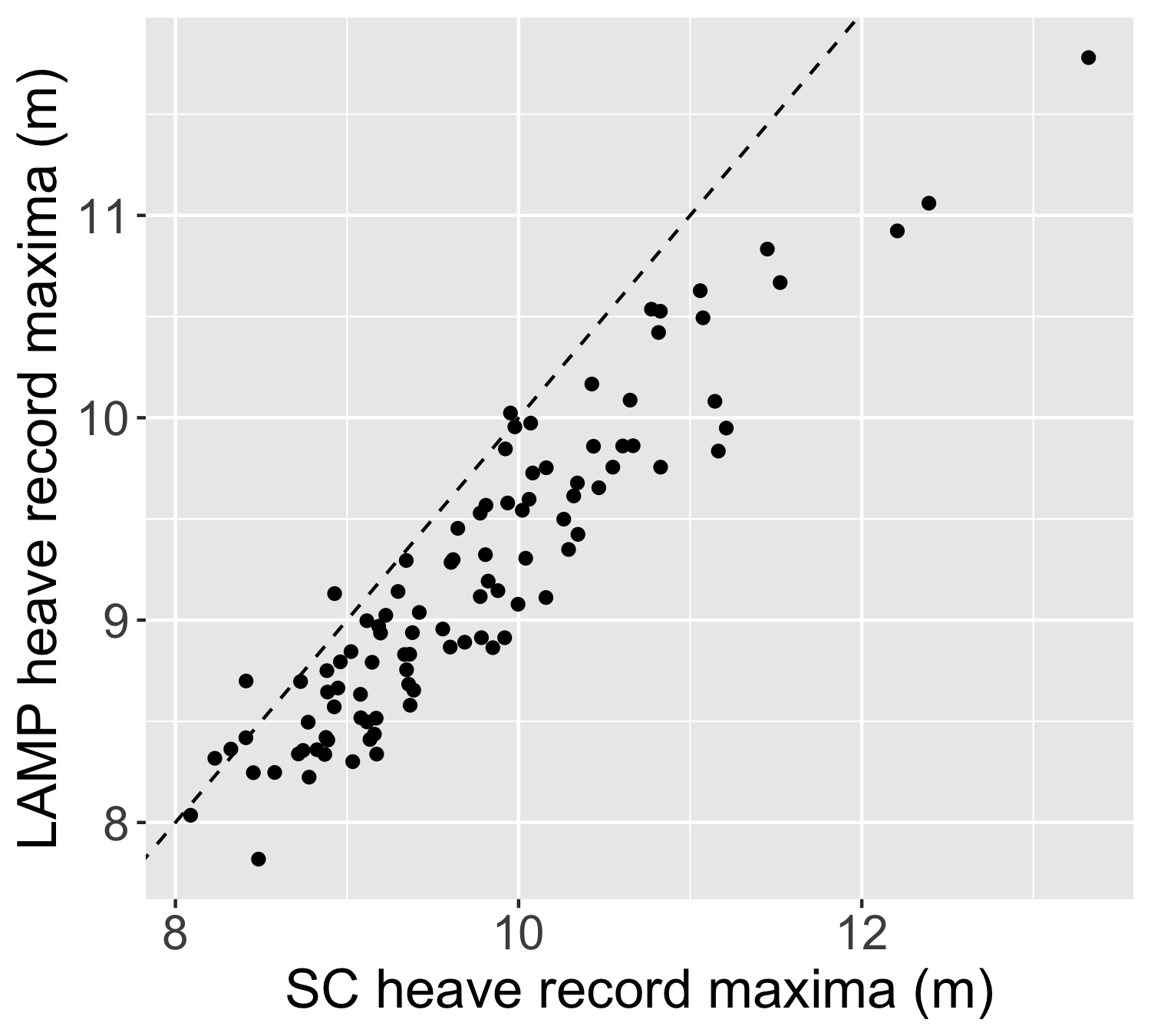}
\includegraphics[width=2.1in,height=1.8in]{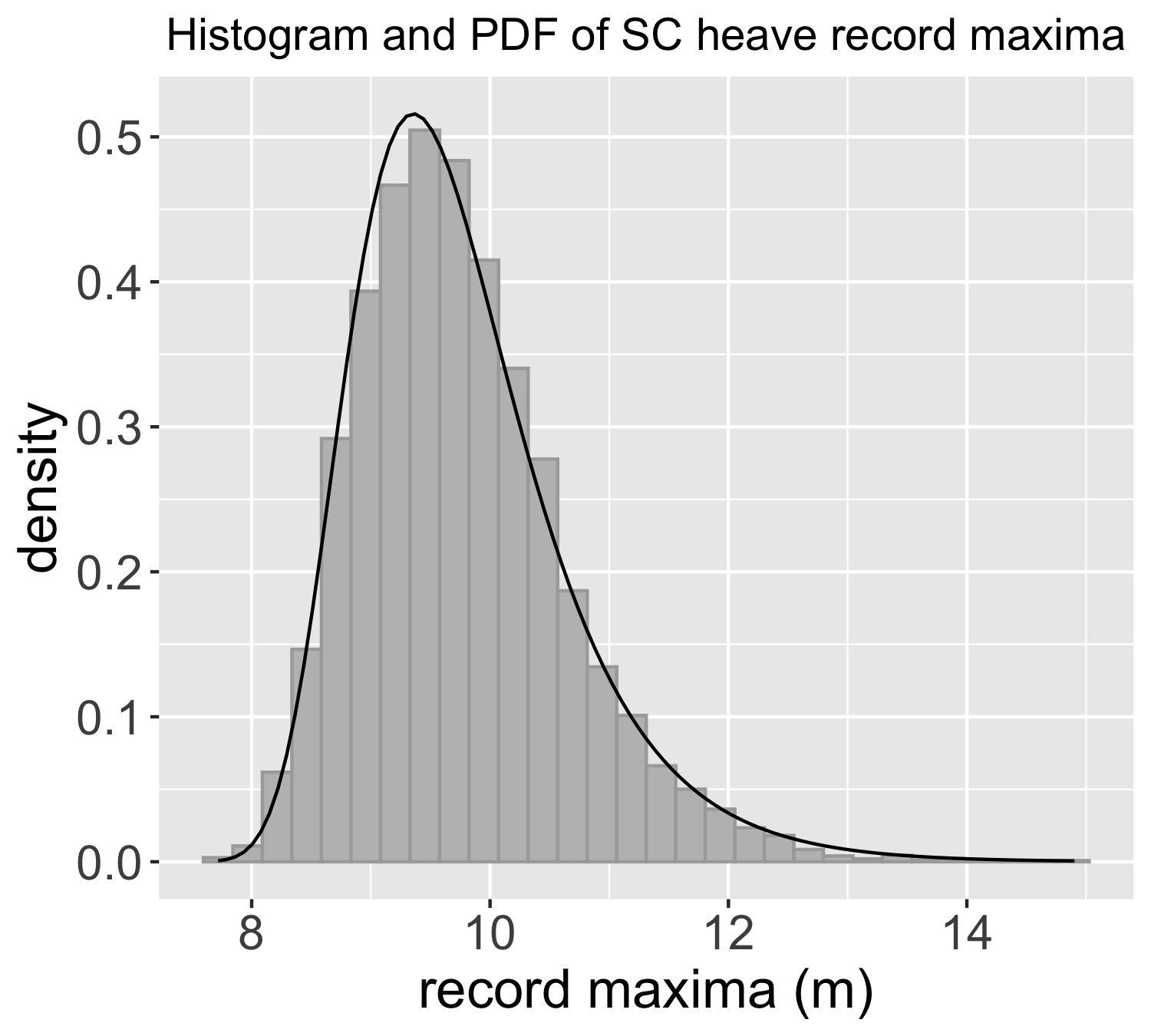}
\includegraphics[width=2.1in,height=1.8in]{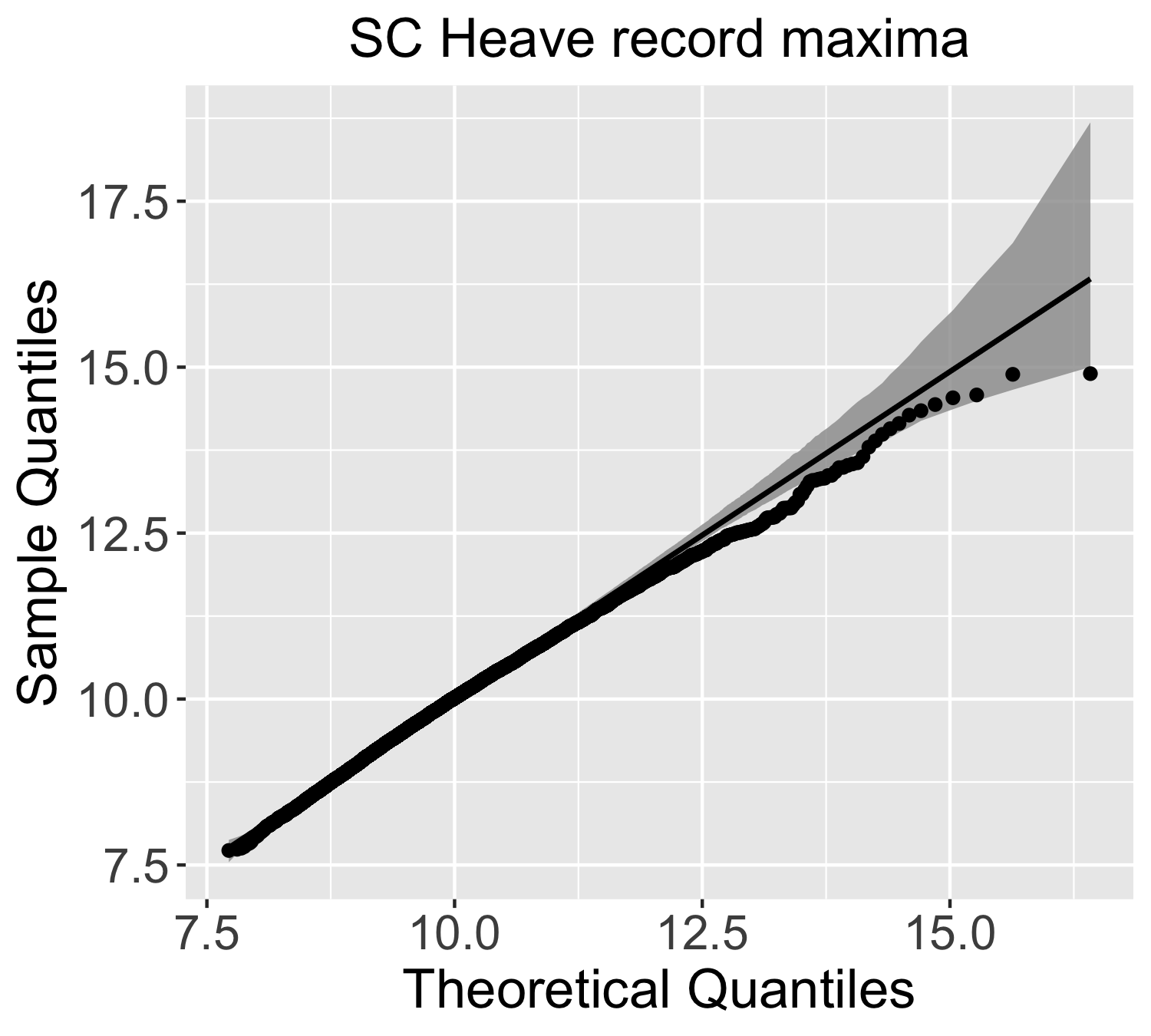}
}
\caption{Left: Scatterplot of record maxima from LAMP and SC. The dashed line is the $45^\circ$ line. Middle: Histogram and estimated p.d.f.\ of SC heave record maxima based on 10,000 observations. Right: QQ plot for SC heave record maxima compared with the estimated Gumbel distribution.}	
\label{f:scatter}
\end{figure}
The MF dataset \eqref{e:data:mf} is then obtained by collecting i.i.d.\ pairs of $(\yhifi,\ylofi)$ for $n=100$ records. 
Additionally, we collect $m=10^4-100$ low-fidelity record maxima, assuming $n+m= 10^4$ is sufficiently large to consider the low-fidelity true parameters known when comparing asymptotic variances.
\Cref{f:scatter}, left plot, presents the scatterplot of the pairs $(\yhifi, \ylofi)$ of heave maxima for SC and LAMP obtained for 100 joint observations.
The middle plot of \Cref{f:scatter} shows the histogram and estimated
density of SC heave record maxima $\ylofi_i, ~i=1, \dots, n+m$.
This p.d.f.\ is modeled by a Gumbel distribution with parameters estimated via ML. 
\Cref{f:scatter}, right plot, shows the QQ plot for SC heave record maxima compared to the estimated Gumbel distribution.
We model the joint distribution with the bivariate Gumbel distribution using \eqref{e:blockmax-G} and \eqref{e:blockmax-G-log}.

\begin{figure}[t]
\centering
\begin{subfigure}{.49\linewidth}
\includegraphics[width=1\linewidth]{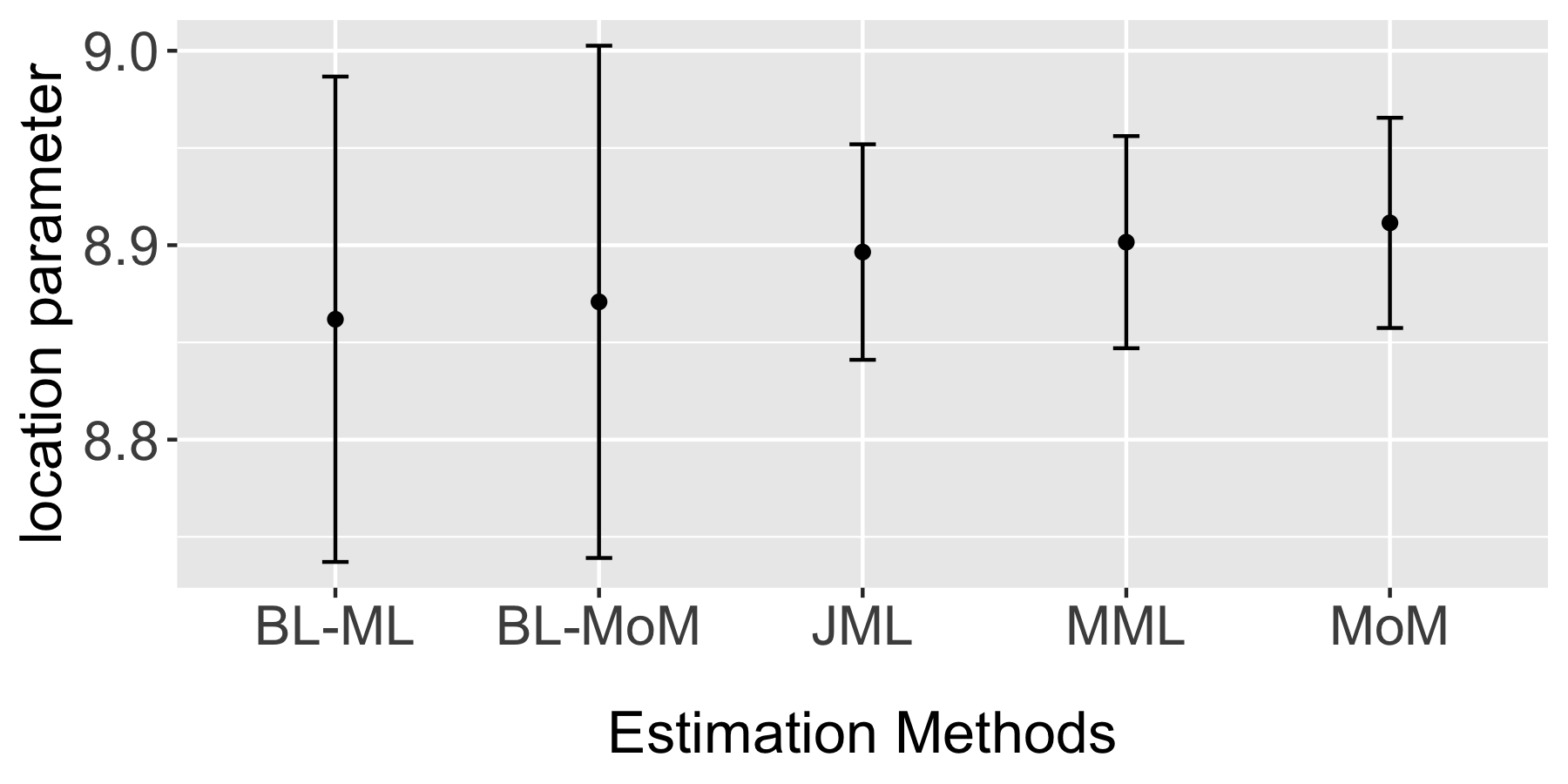}
    \vspace{-2.5em}
\end{subfigure}
\begin{subfigure}{.49\linewidth}
\includegraphics[width=1\linewidth]{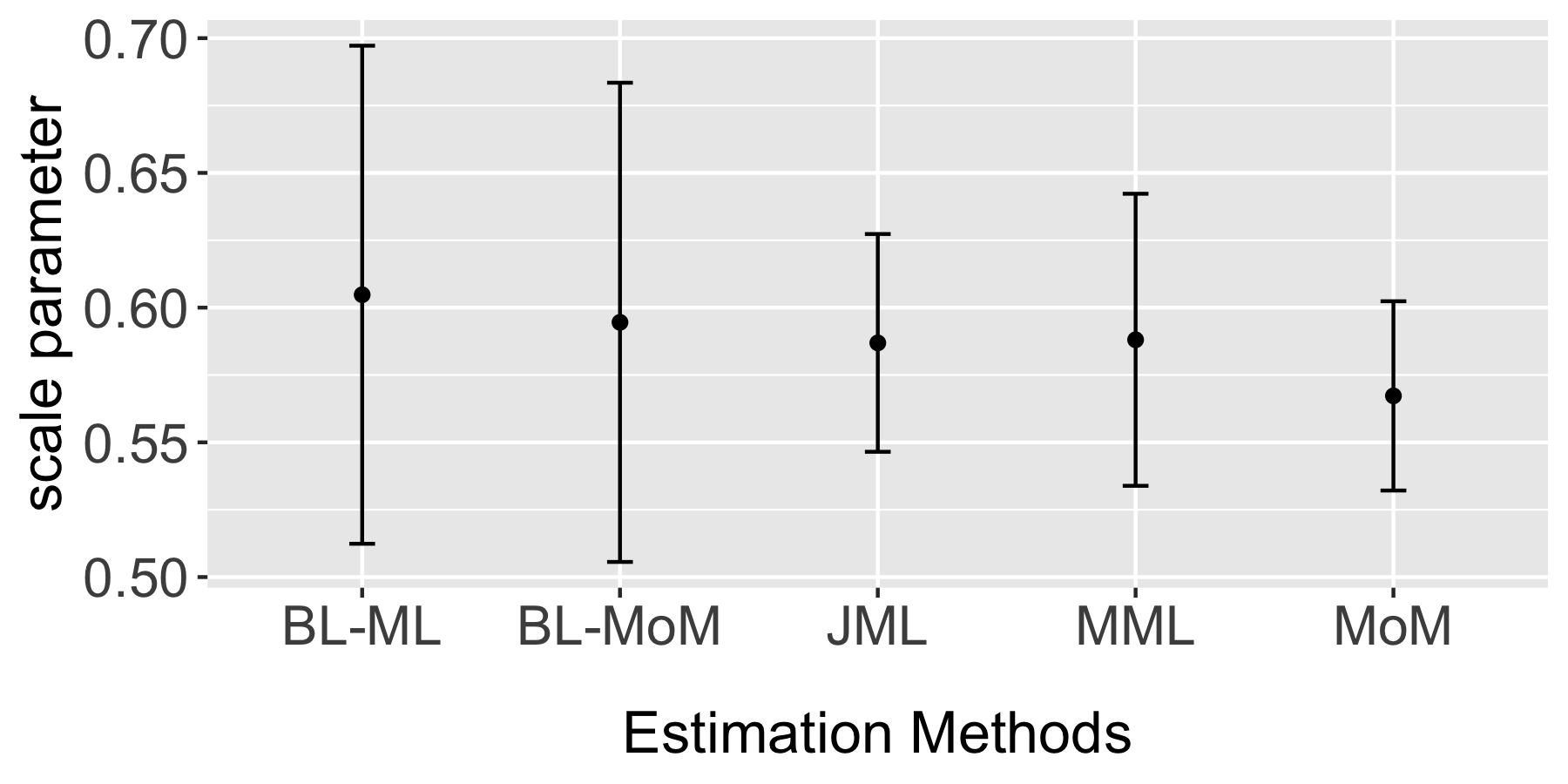}
     \vspace{-2.5em}
\end{subfigure}
\caption{Parametric baseline and MF methods for estimating location (top, $\mu_1$) and scale (bottom, $\sigma_1$) parameters of the Gumbel distribution fitted to heave motion data. The results show baseline and MF estimators along with their confidence intervals.}	
\label{f:heave-par}
\vspace{2em}
\centering
\begin{subfigure}{.49\linewidth}
\includegraphics[width=1\linewidth]{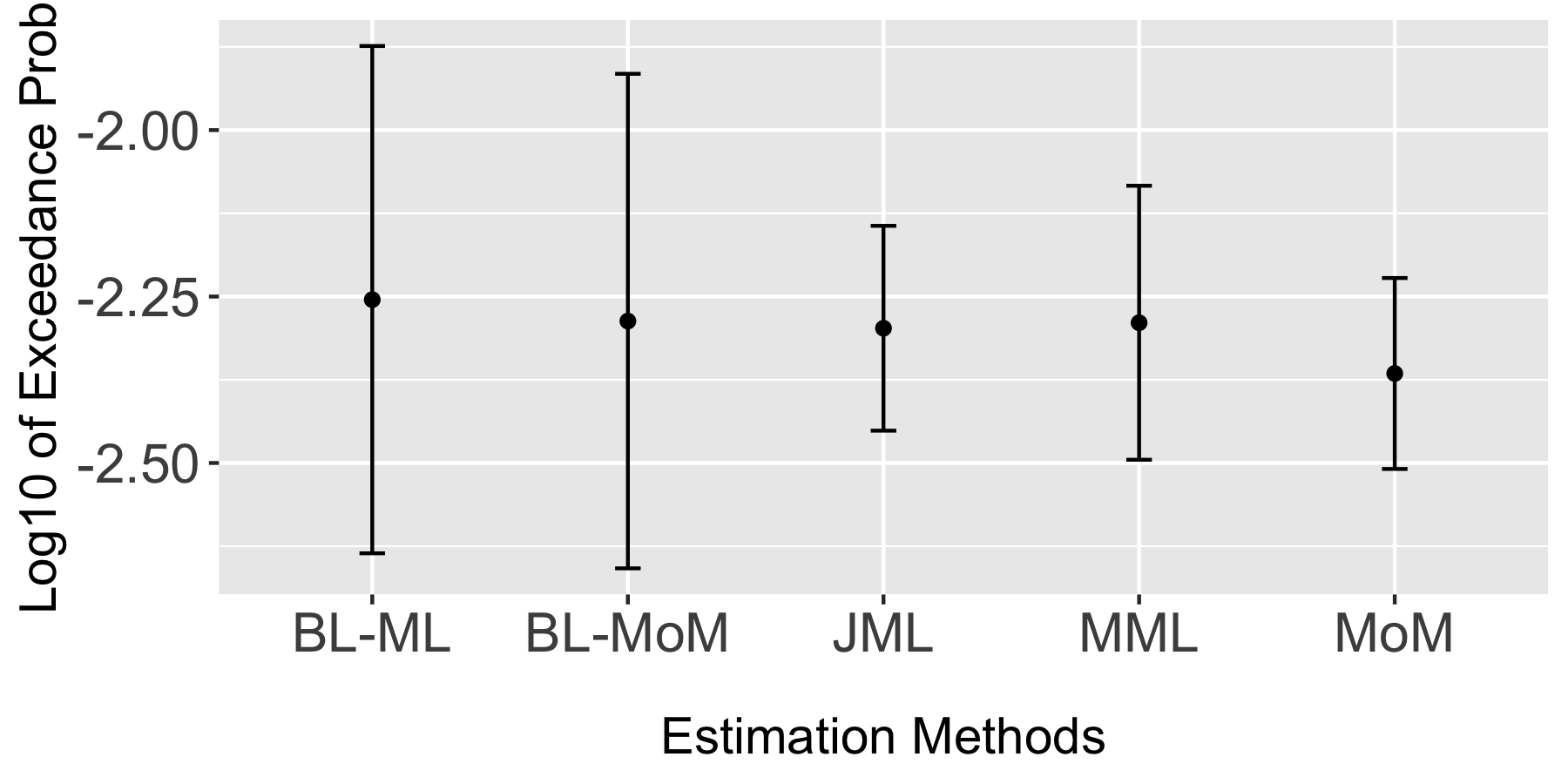}
    \vspace{-2.5em}
\end{subfigure}
\begin{subfigure}{.49\linewidth}
\includegraphics[width=1\linewidth]{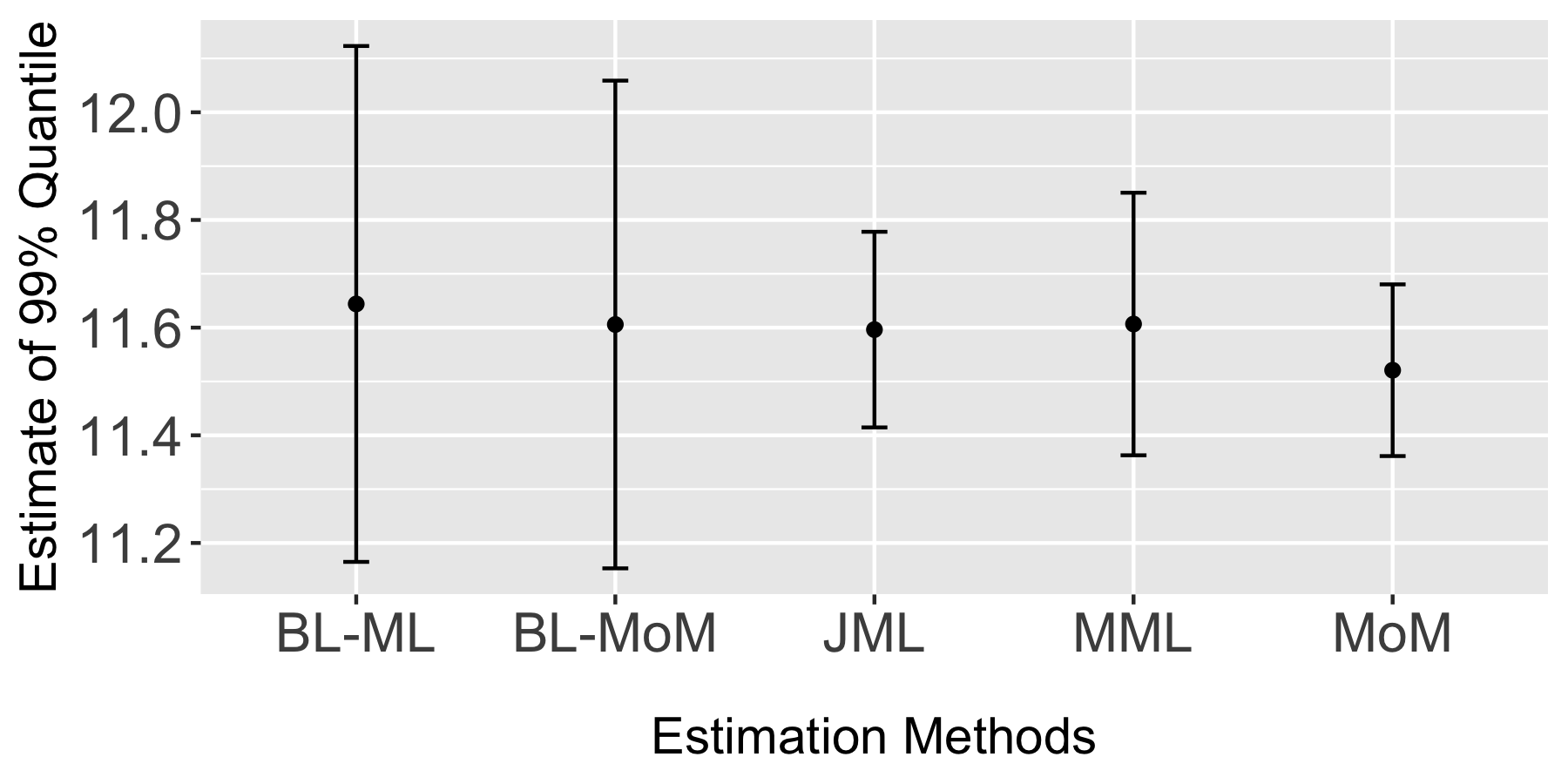}
     \vspace{-2.5em}
\end{subfigure}
\caption{Parametric baseline and MF methods for estimating log10 exceedance probability with $a_1=12$ (top) and a quantile with $p_1 = 0.99$ (bottom) with confidence intervals.}	
\label{f:heave-QoI}
\end{figure}

Applying parametric MF methods discussed in \cref{s:approach},
\Cref{f:heave-par} depicts the resulting baseline (ML and moment) and MF (JML, MoM, and MML) estimators of the high-fidelity parameters estimated along with their confidence intervals. 
The confidence intervals were calculated based on the asymptotic normal distribution of the parameters obtained through ML and moment estimators.
Overall, the MF methods produce narrower confidence intervals than their respective baselines for both the location and scale parameters, indicating improved estimation efficiency.
For the location parameter, the three MF methods yield comparable variances, whereas for the scale parameter, JML achieves the smallest variance, followed by MoM and then MML.
The observed variance behavior is consistent with the results in \cref{s:ex-bvgumbel} and \Cref{f:bvg-unknownS}, where such benefits were seen under strong dependence.
As seen from \Cref{f:scatter}, left plot, the dependence between $\yhifi$ and $\ylofi$ is particularly strong and the advantages of MF estimation are clear.

Finally, \Cref{f:heave-QoI} presents the performance of the parametric baseline and MF methods for estimating extreme QoIs. Specifically, it considers the log10 of exceedance probability with $ a_1=12$ and a quantile with $p_1=0.99$, along with confidence intervals, as discussed in \cref{ex:qoi,ex:qoi-gumbel}.
For both quantities, the advantages of MF estimation are apparent, with JML and MoM yielding the smallest uncertainty, followed by MML.
This is consistent with parameter estimation results in \Cref{f:heave-par}, where the scale parameter in particular showed lower variance for JML and MoM.
It is also worth emphasizing that these quantities are particularly challenging to estimate from the high-fidelity data alone.
Specifically, the maximum observation of the high-fidelity data was 11.78, thus contained no observations exceeding $a_1$. See also \Cref{f:scatter}, left plot. Similarly, with only $n=100$ observations, the direct information from the high-fidelity data about the $p_1=0.99$ quantile is limited. 
We circumvent these limitations by using the available data to fit parametric distributions and viewing these quantities as functions of the distribution parameters as specified in \eqref{e:qoi-Gumbel-exceedP} and \eqref{e:qoi-Gumbel-quantile}. This allows us to calculate confidence intervals by treating these estimates as asymptotically normal, using the delta method, with their variance given by \eqref{e:qoi-avar}.

\section{Conclusions}
\label{s:conclusions}
This work investigated efficient parameter estimation methods for fitting parametric distributions in an MF setting, where both high- and low-fidelity data are available. (Closely related settings are often studied as semi-supervised learning in the machine learning/statistics literature.)
We introduced and evaluated three multi-fidelity parameter estimation techniques: joint maximum likelihood, moment multi-fidelity, and marginal maximum likelihood multi-fidelity estimation. These methods were applied to several parametric models, demonstrating that leveraging the low-fidelity data can reduce the uncertainty in parameter estimation, especially when there is a strong dependence between the high- and low-fidelity data.

An application involving extreme ship motions illustrated the practical advantages of our approach, particularly in extreme value analysis.
By casting the estimation of quantities of interest as a problem of fitting parametric distributions, our approach extends MFMC methods to estimate extremal quantities in the scenarios where the high-fidelity data do not allow direct estimation and extrapolating distributions need to be used.
A future direction is to integrate these methods into cost-aware allocation strategies that balance variance reduction with computational budget,  as briefly discussed in \cref{s:discussion:mvr}.

\section{Disclosure statement}\label{disclosure-statement}

The authors report that there are no competing interests to declare.

\newpage
\bigskip
\begin{center}
{\large\bf SUPPLEMENTARY MATERIAL}
\end{center}

\begin{description}
\item[Reproducible R code:]
R scripts used to reproduce the simulations and real data analysis presented in the manuscript. The code includes all necessary functions and data-loading steps for replicability. (.zip file)

\item[Supplementary Material:]
A separate document containing technical details and additional proofs of key results stated in the main text. (.pdf file)
\end{description}

\phantomsection\label{supplementary-material}
\bigskip

This document provides supplementary material for the paper ``Parametric Multi-Fidelity Monte Carlo Estimation With Applications to Extremes.'' It contains technical details and proofs of key results stated in the main text.

\appendix
\numberwithin{equation}{section}
\section{Optimal coefficients for the MoM estimator when $d_1=2$}
\label{appendix:mom}
In this section, we provide the optimal coefficients for the MoM estimator when $d_1=2$, which are discussed in the main paper following \eqref{e:mfmc-mom-alpha-l}.
We introduce the following notation:
\begin{align*}
V_{H,i} &= \var(\zhifi_i),~V_{L,i} = \var(\zlofi_i) ~\text{for } i=1,2,\\
C_{HL,ij} &= \cov(\zhifi_i, \zlofi_j), ~
C_{HH,ij} = \cov(\zhifi_i, \zhifi_j),~
C_{LL,ij} = \cov(\zlofi_i, \zlofi_j) ~\text{for } i,j=1,2,
\end{align*}
where $H$ stands for high-fidelity, $L$ stands for low-fidelity, and $Z_i^{(j)} = h_i(\yjifi)$ for $i,j=1,2$.
\begin{proposition}
\label{prop:mom}
Consider the MoM estimator as defined in \eqref{e:mom-first}. Assume that $\bfG = [\bfG_1, \bfG_2]^\top \in \bbR^{2\times 2}$ (or its sample analogue $\widehat{\bfG}$) does not depend on $\balpha$, and that $\corr(\zlofi_1,\zlofi_2)\neq 1$.
Then the optimal coefficients $\balpha_k = (\alpha_{k,1}, \alpha_{k,2})^\top$, $k=1,2$, defined in \eqref{e:mfmc-mom-alpha-l}, are given by:
\begin{align*}
\alpha_{k,1} &= \frac{V_{L,2}(g_{k1} C_{HL,11}
+g_{k2} C_{HL,21}) - C_{LL,12} (g_{k2} C_{HL,22} + g_{k1} C_{HL,12}) }{g_{k1} (V_{L,1}V_{L,2} - C_{LL,12}^2) }, \quad \text{provided } g_{k1} \neq 0,\\
\alpha_{k,2} &= \frac{V_{L,1}(g_{k2} C_{HL,22} + g_{k1} C_{HL,12}) - C_{LL,12}(g_{k1} C_{HL,11}
+g_{k2} C_{HL,21}) }{g_{k2} (V_{L,1}V_{L,2} - C_{LL,12}^2)}, \quad \text{provided } g_{k2} \neq 0.
\end{align*}
\end{proposition}

\begin{proof}
Let $ \widetilde{\var} (\htheta_{1,{\rm mom}}) = \bfG\var(\widehat\mu_Z) \bfG^\top$, as implied by \eqref{e:mfmc-mom-var-tilde}.
We first aim to find the optimal coefficients $\balpha_1$ by minimizing $ \widetilde{\var} ((\htheta_{1,{\rm mom}})_1)$. By using \eqref{e:mfmc-moment-Cov-mu}, we have
\begin{align*}
\frac{\partial \widetilde{\var} ((\htheta_{1,{\rm mom}})_1)}{\partial \alpha_{1,1}}
= \frac{2g_{11}m}{n(n+m)}\big\{ &
g_{11} V_{L,1}\alpha_{1,1} +g_{12}C_{LL,12}\alpha_{1,2}  -g_{11}C_{HL,11}-g_{12}C_{HL,21}
\big\},\\
\frac{\partial \widetilde{\var} ((\htheta_{1,{\rm mom}})_1)}{\partial \alpha_{1,2}} = \frac{2g_{12}m}{n(n+m)}\big\{ &g_{11}C_{LL,12} \alpha_{1,1} + g_{12}V_{L,2}\alpha_{1,2} -g_{12}C_{HL,22} -g_{11}C_{HL,12} \big\}.
\end{align*}
Solving $\partial_{\balpha_1}\widetilde{\var}\big((\htheta_{1,{\rm mom}})_1\big) = 0$ leads to the following set of equations:
\begin{align}
\label{e:app-mom-v1a1}
&g_{11} V_{L,1}{\alpha_{1,1}}
+g_{12}C_{LL,12}{\alpha_{1,2}}
=g_{11}C_{HL,11}
+g_{12}C_{HL,21}, \\
\label{e:app-mom-v1a2}
&g_{11}C_{LL,12}{\alpha_{1,1}}
+g_{12}V_{L,2}{\alpha_{1,2}}
=g_{12}C_{HL,22}
+g_{11}C_{HL,12}.
\end{align}
To solve for $\alpha_{1,1}$, we multiply \eqref{e:app-mom-v1a1} by $V_{L,2}$ and subtract $C_{LL,12}$ times \eqref{e:app-mom-v1a2}:
\begin{align*}
g_{11} (V_{L,1}V_{L,2} - C_{LL,12}^2)  \alpha_{1,1} = V_{L,2}(g_{11}C_{HL,11} + g_{12}C_{HL,21}) - C_{LL,12}(g_{12}C_{HL,22}+g_{11}C_{HL,12}).
\end{align*}
Similarly, to solve for $\alpha_{1,2}$, we multiply \eqref{e:app-mom-v1a2} by $V_{L,1}$ and subtract $C_{LL,12}$ times \eqref{e:app-mom-v1a1}:
\begin{align*}
g_{12} (V_{L,1}V_{L,2} - C_{LL,12}^2) \alpha_{1,2} =  V_{L,1}(g_{12}C_{HL,22} + g_{11}C_{HL,12}) - C_{LL,12}(g_{11}C_{HL,11}+g_{12}C_{HL,21}).
\end{align*}
The derivation for $\balpha_2$ follows similarly by considering \( \widetilde{\var} ((\htheta_{1,{\rm mom}})_2) \).
The proposition follows from the fact that $V_{L,1}V_{L,2} - C_{LL,12}^2=0 $ is equivalent to $ \corr(\zlofi_1,\zlofi_2)=1$.
\end{proof}

\begin{example}
\label{appen-ex-mom}
\Cref{prop:mom} implies \eqref{e:bvn-m1-mom-alpha} for the bivariate Gaussian case. Indeed, in that case, by \eqref{e:bvn-mom-form}, we have
\begin{align*}
&\bfG = \begin{pmatrix}
    1 & 0\\
    -2\mu_{1}^\ast & 1
\end{pmatrix}, ~V_{L,1} = \var(\zlofi_1) = (\sigma_2^\ast)^2, ~V_{L,2} = \var(\zlofi_2) = 2(\sigma_2^\ast)^4 + 4 (\mu_2^\ast)^2(\sigma_2^\ast)^2, \\
&C_{HL,11} = \sigstarh\sigstarl\rhostar, ~ C_{HL,22} = 2(\sigstarh)^2(\sigstarl)^2(\rhostar)^2 + 4 \mustarh\mustarl\sigstarh\sigstarl\rhostar, \\
&C_{HL,12} = 2\mustarl\sigstarh\sigstarl\rhostar, ~ C_{HL,21} = 2\mustarh\sigstarh\sigstarl\rhostar, ~
C_{LL,12} = 2\mustarl(\sigstarl)^2.
\end{align*}
Since $g_{12} = 0$ and $g_{11}=1$, equation \eqref{e:app-mom-v1a1} simplifies to $V_{L,1} \alpha_1 = C_{HL,11}$, yielding
\begin{align*}
\alpha_1 = \frac{C_{HL,11}}{V_{L,1}}= \rhostar \frac{\sigstarh}{\sigstarl},
\end{align*}
leading to $\alpha_{1, {\rm opt}}$ in \eqref{e:bvn-m1-mom-alpha}. For $\balpha_{2, {\rm opt}}$, \cref{prop:mom} gives
\begin{align*}
\alpha_{2,1} & = \frac{- 2\mustarl(\sigstarl)^2\left\{
2(\sigstarh)^2(\sigstarl)^2(\rhostar)^2 + 4 \mustarh\mustarl\sigstarh\sigstarl\rhostar -4\mustarh\mustarl\sigstarh\sigstarl\rhostar
\right\} }{-2\mustarh\left\{2(\sigma_2^\ast)^6 + 4 (\mu_2^\ast)^2(\sigma_2^\ast)^4 - 4(\mustarl)^2(\sigstarl)^4\right\}} = \frac{\mustarl}{\mustarh} \left(\rhostar\frac{\sigstarh}{\sigstarl}\right)^2,\\
\alpha_{2,2} & = \frac{(\sigstarl)^2\left\{
2(\sigstarh)^2(\sigstarl)^2(\rhostar)^2 + 4 \mustarh\mustarl\sigstarh\sigstarl\rhostar -4\mustarh\mustarl\sigstarh\sigstarl\rhostar
\right\}}{2(\sigstarl)^6} = \left(\rhostar\frac{\sigstarh}{\sigstarl}\right)^2,
\end{align*}
as stated.
\end{example}

\section{Proof of Proposition~\ref{prop:bvn:joint-mle}}
\label{appendix:bvn}

We consider the bivariate Gaussian distribution
\begin{gather}
\label{e:proof-bvn}
   \bfY :=  \left( \begin{array}{c}
         \yhifi  \\
         \ylofi
    \end{array} \right) \sim \calN\left( \bmu, \bSigma  \right), \quad \bSigma = \begin{pmatrix}
         \sigma_1^2 & \rho\sigma_1\sigma_2 \\
         \rho\sigma_1\sigma_2 & \sigma_2^2
    \end{pmatrix}.
\end{gather}
We introduce the following notation: $\bfY_i=(\yhifi_i, \ylofi_i)^\top$ denotes the $i$th observation, where $i=1, \dots, n+m$; $\Bar \bfY_{a:b}=\frac{1}{b-a+1}\sum_{i=a}^b \bfY_i$ denotes the sample mean of the data; $\bfe_1=(1,0)^\top, \bfe_2=(0,1)^\top$ denote unit vectors; and for a matrix $\bfA$, $|\bfA|$ denotes its determinant.
The negative log-likelihood of the joint distribution \eqref{e:proof-bvn} is given by
\begin{equation}
    \begin{split}
    \label{e:app-nll}
        \ell(\bmu, \bSigma)
        &= \frac{n}{2}\log|\bSigma|  + \frac{1}{2}\sum_{i=1}^{n} (\bfY_i - \bmu)^\top\bSigma^{-1}(\bfY_i-\bmu) \\
        &\quad + \frac{m}{2}\log|\bfe_2^\top \bSigma \bfe_2|
        + \frac{1}{2}\sum_{i=n+1}^{n+m}  (\bfY_{i} - \bmu)^\top \bfe_2 (\bfe_2^\top\bSigma \bfe_2)^{-1} \bfe_2^\top (\bfY_{i}-\bmu).
    \end{split}
\end{equation}

To find the maximum likelihood estimators, we first solve the likelihood equation with respect to $\bmu$. Setting ${\partial \ell}/{\partial \bmu} = 0$ yields
$$
n \bSigma^{-1}(\Bar \bfY_{1:n}-\bmu)
        +m (\bfe_2^\top\bSigma \bfe_2)^{-1} \bfe_2  \bfe_2^\top  (\Bar\bfY_{n+1:n+m}-\bmu) =0.
$$
Solving for $\bmu$, we obtain
\begin{equation}
    \begin{split}
\label{e:proof-mu-first}
\widehat \bmu  = &  ~ ( n \bSigma^{-1} +m  (\bfe_2^\top\bSigma \bfe_2)^{-1} \bfe_2 \bfe_2^\top )^{-1} (n \bSigma^{-1}\Bar \bfY_{1:n} + m (\bfe_2^\top\bSigma \bfe_2)^{-1} \bfe_2 \bfe_2^\top  \Bar\bfY_{n+1:n+m}).
    \end{split}
\end{equation}

We use the Sherman--Morrison-type identity: 
\begin{align}
\label{e:app-lemma}
(\bfB+\bfc\bfc^\top)^{-1} = \bfB^{-1} - \frac{\bfB^{-1}\bfc\bfc^\top \bfB^{-1}}{1+\bfc^{\top}\bfB^{-1}\bfc},
\end{align}
where $\bfc$ is a vector and $\bfB$ and $\bfB+\bfc\bfc^\top$ are nonsingular square matrices.
Then,
\begin{align}
( n \bSigma^{-1} +m (\bfe_2^\top\bSigma \bfe_2)^{-1} \cdot \bfe_2 \bfe_2^\top )^{-1}& =
\frac{1}{n}\bSigma -\frac{m}{n^2}(\bfe_2^\top\bSigma \bfe_2)^{-1} \frac{\bSigma \bfe_2\bfe_2^\top \bSigma}{1+ \frac{m}{n}(\bfe_2^\top\bSigma \bfe_2)^{-1}  \bfe_2^{\top}\bSigma \bfe_2}\notag \\
& =
\frac{1}{n}\bSigma -\frac{m}{n(n+m)} (\bfe_2^\top\bSigma \bfe_2)^{-1}{\bSigma \bfe_2\bfe_2^\top \bSigma}.
\end{align}
Substituting this into \eqref{e:proof-mu-first},
\begin{equation}
\begin{split}
\widehat \bmu
&=  \BYn  -
 \frac{m}{n+m} (\bfe_2^\top\bSigma \bfe_2)^{-1}\bSigma \bfe_2\bfe_2^\top \BYn   + \frac{m}{n+m}(\bfe_2^\top\bSigma \bfe_2)^{-1} \bSigma \bfe_2\bfe_2^\top \BYm \\
\label{e:proof-mu-final}
&=\BYn + (\bfe_2^\top\bSigma \bfe_2)^{-1}\bSigma \bfe_2\bfe_2^\top ( \BYnm - \BYn).
\end{split}
\end{equation}
To express the solution elementwise, we write
\begin{align}
\label{e:app-elem-mu}
\widehat\mu_2 =  \bfe_2^\top \widehat{\bmu} &= \bfe_2^\top \Bar \bfY_{1:n} + \bfe_2^\top ( \BYnm - \BYn) = \bfe_2^\top \Bar\bfY_{1:n+m} = \ybart,\\
\widehat\mu_1 = \bfe_1^\top \widehat{\bmu} &= \bfe_1^\top \Bar \bfY_{1:n}  + \frac{(\bfe_1^\top\bSigma \bfe_2)}{(\bfe_2^\top\bSigma \bfe_2)}\bfe_2^\top ( \BYnm - \BYn) = \ybaro + \rho\frac{\sigma_1}{\sigma_2}\left(\ybart-\ybarto\right).\notag
\end{align}

Next, we solve the likelihood equation with respect to $\bSigma$, using the optimal $\hmu$ in \eqref{e:proof-mu-final}. We first rewrite the sum of squared deviations in the log-likelihood function \eqref{e:app-nll} as
\begin{align*}
\sum_{i=1}^{n} (\bfY_i - \bmu)^\top\bSigma^{-1}(\bfY_i-\bmu) =\sum_{i=1}^{n} (\bfY_i -\BYn )^\top\bSigma^{-1}(\bfY_i-\BYn) + n (\bmu-\BYn )^\top\bSigma^{-1}(\bmu-\BYn) .
\end{align*}
Substituting \eqref{e:proof-mu-final}, we have
\begin{align*}
(\hmu-\BYn )^\top\bSigma^{-1}(\hmu-\BYn) &=  \frac{1}{\bfe_2^\top\bSigma \bfe_2} (\BYnm - \BYn)^\top \bfe_2 \bfe_2^\top (\BYnm - \BYn) =: \frac{1}{\bfe_2^\top\bSigma \bfe_2} V_n,
\end{align*}
where $$V_n = (\BYnm - \BYn)^\top \bfe_2 \bfe_2^\top (\BYnm - \BYn) = \left(\ybart -\ybarto\right)^2.$$ To differentiate the function $\ell$ with respect to $\bSigma$, we use the following facts:
$$
\text{tr}\left(\sum_{i=1}^{n} (\bfY_i - \BYn)^\top\bSigma^{-1}(\bfY_i-\BYn)\right) = \text{tr}\left(\bSigma^{-1}\bfS_n\right),$$
$$
\frac{\partial \log|\bSigma|}{\partial \bSigma} = (\bSigma^{-1})^\top, \quad
\frac{\partial \text{tr}(\bSigma^{-1}\bfS_n)}{\partial \bSigma} = -(\bSigma^{-1}\bfS_n\bSigma^{-1})^\top,$$
where $$\bfS_n = \sum_{i=1}^{n} (\bfY_i - \BYn)(\bfY_i - \BYn)^\top.$$
Then, the likelihood equation with respect to $\bSigma$, i.e., ${\partial \ell(\bmu, \bSigma)}/{\partial \bSigma}=0$ is
\begin{equation} \label{e:app-nll-sig}
\begin{split}
\frac{n}{2} \boldsymbol{\Sigma}^{-1} - \frac{1}{2} \boldsymbol{\Sigma}^{-1}\mathbf{S}_n\boldsymbol{\Sigma}^{-1}
+ \frac{1}{2}\left[ \frac{m}{\mathbf{e}_2^\top\boldsymbol{\Sigma} \mathbf{e}_2} - \frac{n V_n + T_m}{(\mathbf{e}_2^\top\boldsymbol{\Sigma} \mathbf{e}_2)^{2}} \right] \mathbf{e}_2 \mathbf{e}_2^\top = 0,
\end{split}
\end{equation}
where $$T_m = \sum_{i=n+1}^{n+m} \bfe_2^\top (\bfY_{i} - \widehat\bmu)(\bfY_{i} - \widehat\bmu)^\top \bfe_2 = \sum_{i=n+1}^{n+m} (\ylofi_i - \ybart)^2.$$

Multiplying both sides of \eqref{e:app-nll-sig} by $\bSigma$ and rearranging terms, the solution $\widehat\bSigma$ should satisfy
\begin{align}
\label{e:proof-sig}
n \widehat{\mathbf{\Sigma}} &= \mathbf{S}_n
 + \left[ \frac{n V_n + T_m}{(\mathbf{e}_2^\top\widehat{\mathbf{\Sigma}} \mathbf{e}_2)^{2}} - \frac{m}{\mathbf{e}_2^\top\widehat{\mathbf{\Sigma}} \mathbf{e}_2} \right] \widehat{\mathbf{\Sigma}} \mathbf{e}_2 \mathbf{e}_2^\top \widehat{\mathbf{\Sigma}}.
\end{align}
We are now able to obtain elementwise solutions as follows. For the variance component $\widehat\sigma_2^2 = \mathbf{e}_2^\top \widehat{\boldsymbol{\Sigma}} \mathbf{e}_2$, the equation \eqref{e:proof-sig} implies
\begin{align}
\label{e:proof-s22-def}
\bfe_2^\top \hSig \bfe_2 &= \frac{1}{n}\bfe_2^\top \bfS_n \bfe_2 + V_n + \frac{1}{n}T_m  - \frac{m}{n} \bfe_2^\top \hSig \bfe_2.
\end{align}
Thus, we get
\begin{align}
{{\widehat\sigma_2}^2} &= \bfe_2^\top \hSig \bfe_2 = \frac{1}{n+m}\bfe_2^\top \bfS_n \bfe_2 + \frac{n}{n+m}V_n  + \frac{1}{n+m}T_m\notag\\
&= \frac{1}{n+m} \left\{\sum_{i=1}^n (\ylofi_i-\ybarto)^2 + n \left(\ybart - \ybarto\right)^2 +  \sum_{i=n+1}^{n+m} (\ylofi_{i}-\ybart)^2\right\}\notag\\
\label{e:proof-s22}
&= \frac{1}{n+m} \sum_{i=1}^{n+m} (\ylofi_i-\ybart)^2.
\end{align}
Similarly, from \eqref{e:proof-sig},
\begin{align}
\bfe_1^\top \hSig \bfe_2 &= \frac{1}{n}\bfe_1^\top \bfS_n \bfe_2 + \frac{\bfe_1^\top\hSig \bfe_2}{\bfe_2^\top\hSig \bfe_2} \left\{V_n + \frac{1}{n}T_m  - \frac{m}{n} \bfe_2^\top \hSig \bfe_2\right\}\notag\\
\label{e:app-s12-help}
&= \frac{1}{n}\bfe_1^\top \bfS_n \bfe_2 + \frac{\bfe_1^\top\hSig \bfe_2}{\bfe_2^\top\hSig \bfe_2} \left\{\bfe_2^\top \hSig \bfe_2 - \frac{1}{n}\bfe_2^\top \bfS_n \bfe_2 \right\},
\end{align}
where the second equality follows from \eqref{e:proof-s22-def}. Thus, we get
\begin{align}
\label{e:proof-s12}
\left(\widehat{\rho \frac{\sigma_1}{\sigma_2}}\right)=\frac{\bfe_1^\top \hSig \bfe_2}{\bfe_2^\top \hSig \bfe_2} & = \frac{\bfe_1^\top \bfS_n \bfe_2}{\bfe_2^\top \bfS_n \bfe_2}.
\end{align}
Finally, from \eqref{e:proof-sig}, by arguing as in \eqref{e:app-s12-help},
\begin{align*}
\widehat\sigma_1^2 = \bfe_1^\top \hSig \bfe_1 &= \frac{1}{n}\bfe_1^\top \bfS_n \bfe_1 + \left(\frac{\bfe_1^\top\hSig \bfe_2}{\bfe_2^\top\hSig \bfe_2}\right)^2 \left\{\bfe_2^\top \hSig \bfe_2 - \frac{1}{n}\bfe_2^\top \bfS_n \bfe_2 \right\}.
\end{align*}
Substituting \eqref{e:proof-s22} and \eqref{e:proof-s12}, we get
\begin{align}
\label{e:app-s11}
\widehat\sigma_1^2 &= \frac{1}{n}\bfe_1^\top \bfS_n \bfe_1 + \left(\frac{\bfe_1^\top \bfS_n \bfe_2}{\bfe_2^\top \bfS_n \bfe_2}\right)^2 \left\{ \widehat\sigma_2^2 - \frac{1}{n}\bfe_2^\top \bfS_n \bfe_2 \right\}.
\end{align}
The proposition now follows from \eqref{e:app-elem-mu}, \eqref{e:proof-s22}, \eqref{e:proof-s12} and \eqref{e:app-s11}.

\section{Proof of Proposition~\ref{prop:ber:jml}}
\label{appendix:ber}
Using the joint p.m.f.\ in \eqref{e:ex-ber-pmf}, the log-likelihood of the re-parameterized model can be written for each pair \((y_1, y_2) \in \{0,1\}^2\) as
\begin{equation}
\label{e:ex-ber-ell}
\begin{split}
\ell_{\tilde\bfeta}(y_1,y_2)& = y_1y_2 \log \gamma + y_1(1-y_2)\log (p_1-\gamma) \\
&+ y_2(1-y_1) \log (p_2-\gamma) + (1-y_1)(1-y_2)\log (1-p_1-p_2+\gamma),
\end{split}
\end{equation}
and the marginal log-likelihood for $\ylofi$ can be written as
\begin{align}
\label{e:ex-ber-lp2}
\ell_{p_2}(y_2) = y_2 \log p_2 + (1-y_2)\log (1-p_2).
\end{align}
For the \(n\) paired observations, introduce the four cell counts
\[
n_{ab} = \sum_{i=1}^{n}\bbone\{Y^{(1)}_i=a,\;Y^{(2)}_i=b\},
  \qquad a,b\in\{0,1\},
\]
so that \(\sum_{a,b}n_{ab}=n\). For the additional \(m\) univariate outcomes, define
\[
m_{b} = \sum_{i=n+1}^{n+m}\bbone\{Y^{(2)}_i=b\},
  \qquad b\in\{0,1\},
\]
giving \(\sum_{b} m_{b}=m\).
Then, the log-likelihood for the MF dataset \eqref{e:data:mf} is given as
\begin{align*}
\ell(p_1, p_2, \gamma) &=\sum_{i=1}^n\ell_{\tilde\bfeta}(\yhifi_i,\ylofi_i) + \sum_{i=n+1}^{n+m} \ell_{p_2}(\ylofi_i) \\
&= n_{11}\log \gamma + n_{10}\log (p_1-\gamma)+ n_{01} \log (p_2-\gamma)+ n_{00}\log (1-p_1-p_2+\gamma)\\
&\quad + m_1\log p_2 + m_0\log (1-p_2).
\end{align*}
Differentiating with respect to $\gamma$, $p_1$, and $p_2$ yields the likelihood equations
\begin{align}
\label{e:ber:dell-dg}
\frac{\partial \ell}{\partial\gamma} &: \frac{n_{11}}{\gamma}  -\frac{n_{10}}{p_1-\gamma} - \frac{n_{01}}{p_2-\gamma} +\frac{n_{00}}{1-p_1-p_2+\gamma} = 0,\\
\label{e:ber:dell-dp1}
\frac{\partial \ell}{\partial p_1} &: \frac{n_{10}}{p_1-\gamma} -\frac{n_{00}}{1-p_1-p_2+\gamma}= 0,\\
\label{e:ber:dell-dp2}
\frac{\partial \ell}{\partial p_2} &: \frac{n_{01}}{p_2-\gamma} -\frac{n_{00}}{1-p_1-p_2+\gamma}
+ \frac{m_1}{p_2}-\frac{m_0}{1-p_2}= 0.
\end{align}
We solve this system of equations to obtain the JML estimators. Observe that
\begin{align}
\label{e:ber:appen:obs1}
&\eqref{e:ber:dell-dp1} : \quad  \frac{n_{00}}{1-\widehat p_1-\widehat p_2+\widehat\gamma} = \frac{n_{10}}{\widehat p_1-\widehat\gamma},\\
&\label{e:ber:appen:obs2}
\eqref{e:ber:dell-dg} + \eqref{e:ber:dell-dp1} : \quad \frac{n_{01}}{\widehat p_2-\widehat\gamma} = \frac{n_{11}}{\widehat\gamma} .\end{align}
Substituting \eqref{e:ber:appen:obs1} and \eqref{e:ber:appen:obs2} into \eqref{e:ber:dell-dp2} gives
\begin{align}
\label{e:ber:appen:obs3}
\frac{n_{11}}{\widehat\gamma} - \frac{n_{10}}{\widehat p_1-\widehat\gamma} + \frac{m_1}{\widehat p_2}-\frac{m_0}{1-\widehat p_2}= 0.
\end{align}
Rearranging \eqref{e:ber:appen:obs1} and \eqref{e:ber:appen:obs2} yields
\begin{align}
\label{e:ber:appen:obs4}
\widehat p_1-\widehat\gamma = \frac{n_{10}}{n_{00}+n_{10}} (1-\widehat p_2),\quad \widehat\gamma = \frac{n_{11}}{n_{01}+n_{11}} \widehat p_2.\end{align}
Substituting \eqref{e:ber:appen:obs4} into \eqref{e:ber:appen:obs3} gives
\begin{align}
\label{e:ber:appen:p2}
\widehat p_{2, {\rm jml}} = \frac{n_{00}+n_{11}+m_0}{n+m} &= \ybart.
\end{align}
Then, by \eqref{e:ber:appen:obs4},
\begin{align}
\label{e:ber:appen:gam}
\widehat \gamma_{\rm jml} = \frac{n_{11}}{n_{01}+n_{11}}\ybart =\frac{\ybart}{\ybarto}\cdot\frac{1}{n}\sum_{i=1}^n \yhifi_i\ylofi_i.
\end{align}
Finally, substituting the expressions for $\widehat \gamma_{\rm jml}$ and $\widehat p_{2, {\rm jml}}$ into \eqref{e:ber:appen:obs4} yields
\begin{align}
\label{e:ber:appen:p1}
\widehat p_{1, {\rm jml}} &= \frac{\frac{1}{n}\sum_{i=1}^n \yhifi_i\ylofi_i}{\ybarto}\ybart + \frac{\ybaro - \frac{1}{n}\sum_{i=1}^n \yhifi_i\ylofi_i}{1-\ybarto}\left(1-\ybart\right) \notag\\
&= \ybaro + \frac{\frac{1}{n}\sum_{i=1}^n \yhifi_i\ylofi_i - \ybarto\ybaro}{\ybarto(1-\ybarto)}\left(\ybart-\ybarto
\right).
\end{align}

\section{Bivariate Gumbel case}
\label{appendix:gumbel}
This section complements Section~\ref{s:ex-bvgumbel} of the main paper, which considers the bivariate Gumbel distribution with both $\mu_1$ and $\sigma_1$ unknown. Here, we consider the simplified case where all parameters are known except for the location parameter $\mu_1$. This assumes the case $m=\infty$ in \eqref{e:mfmc-case-m} and also that $\sigma_1^\ast$ and $r^\ast$ are known.
Under the Gumbel distribution, the moment formulation in \eqref{e:prelim-moment} for $\mu_1$ can be obtained as $\mu_1 = \bbE \yhifi - \sigma_1 \gamma $, where $\gamma$ is Euler's constant and $\sigma_1$ is the scale parameter (e.g.\ \cite{coles:2001}).
The various estimators and their asymptotic variances are as follows: for the baseline estimators,
\begin{align}
\label{e:ex-bvg-base-mom}
&\widehat\mu_{1,{\rm bl, mom}} = \ybaro - \sigma_1^\ast \gamma, ~ n\var(\widehat\mu_{1,{\rm bl, mom}}) = \var(\yhifi), \\
\label{e:ex-bvg-base-ml}
&\widehat\mu_{1,{\rm bl, ml}} = \argmax{\mu_{1}} \prod_{i=1}^n f_{(\mu_1, \sigma_1^\ast)}(\yhifi_i), ~ \lim_n n\var(\widehat\mu_{1,{\rm bl, ml}}) = \left.\left( -\bbE \ddl_1(\yhifi)\right)^{-1}\right|_{\mu_1=\mu_1^\ast};\\
\intertext{for the JML estimator,}
\label{e:ex-bvg-joint-mu}
&\widehat{\mu}_{1,{\rm jml}} = \argmax{\mu_1} \prod_{i=1}^{n} f_{(\mu_1, \sigma_1^\ast, \mu_2^\ast, \sigma_2^\ast, r^\ast)}(\yhifi_i, \ylofi_i), \\
\label{e:ex-bvg-joint-var}
&\lim_n n\var(\widehat{\mu}_{1,{\rm jml}}) =\left. \left(-\bbE \frac{\partial^2 \log f_{(\mu_1, \sigma_1^\ast, \mu_2^\ast, \sigma_2^\ast, r^\ast)}(\yhifi, \ylofi)}{\partial \mu_1^2}\right)^{-1} \right|_{\mu_1=\mu_1^\ast};\\
\intertext{for the MoM estimator,}
\label{e:ex-bvg-mom-mu}
& \widehat{\mu}_{1,{\rm mom}} = \ybaro - \sigma_1^\ast\gamma + \alpha_{\rm opt}\left( \bbE \ylofi - \ybarto \right),\\
\label{e:ex-bvg-mom-var}
& n\var(\widehat{\mu}_{1,{\rm mom}}) = \var(\yhifi)\left(1 - \corr(\yhifi, \ylofi)^2\right);\\
\intertext{for the MML estimator,}
\label{e:ex-bvg-mml-mu}
& \widehat{\mu}_{1,{\rm mml}} = (\widehat\mu_{1,{\rm bl,ml}})_n + \beta_{\rm opt}\left( \mu_2^\ast - (\widehat\mu_{2,{\rm bl,ml}})_{n} \right), \\
\label{e:ex-bvg-mml-var}
& \lim_n n\var(\widehat{\mu}_{1,{\rm mml}}) = \left.c_1^2 \var(\dl_1(\yhifi))\left(1 - \corr(\dl_1(\yhifi), \dl_2(\ylofi))^2\right)\right|_{\mu_1=\mu_1^\ast}.
\end{align}
In \eqref{e:ex-bvg-base-ml} and \eqref{e:ex-bvg-mml-var}, the log-likelihood function is utilized, which is defined as
\begin{align*}
\ell_{\mu_j}(y) &= -\frac{y-\mu_j}{\sigma_j^\ast} - \log(\mu_j) - e^{-(y-\mu_j)/\sigma_j^\ast},\\
\dl_{\mu_j}(y) &= \frac{\partial \ell_{\mu_j}(y)}{\partial \mu_j} = \frac{1}{\sigma_j^\ast} - \frac{1}{\sigma_j^\ast}e^{-(y-\mu_j)/\sigma_j^\ast}, \\ \ddl_{\mu_j}(y) &= \frac{\partial^2 \ell_{\mu_j}(y)}{\partial \mu_j^2} = -\frac{1}{(\sigma_j^\ast)^2} e^{-(y-\mu_j)/\sigma_j^\ast},
\end{align*}
and $c_j := \left.\left(-\bbE \ddl_{\mu_j}(X^{(j)})\right)^{-1}\right|_{\mu_j=\mu_j^\ast} = (\sigma_j^\ast)^2$ represents the inverse of the Fisher information for the marginal distribution.
Then, $\alpha_{\rm opt}$ and $\beta_{\rm opt}$ in \eqref{e:ex-bvg-mom-mu} and \eqref{e:ex-bvg-mml-mu} are defined as
\begin{align}
\label{e:bvg-alpha}
\alpha_{\rm opt} = \frac{\cov(\yhifi, \ylofi)}{\var(\ylofi)}, \quad \beta_{\rm opt} = \frac{c_1}{c_2} \frac{\cov(\dl_{\mu_1^\ast}(\yhifi), \dl_{\mu_2^\ast}(\ylofi))}{\var(\dl_{\mu_2^\ast}(\ylofi))}.
\end{align}

\begin{figure}[t]
\centerline{
\includegraphics[width = .48\textwidth]{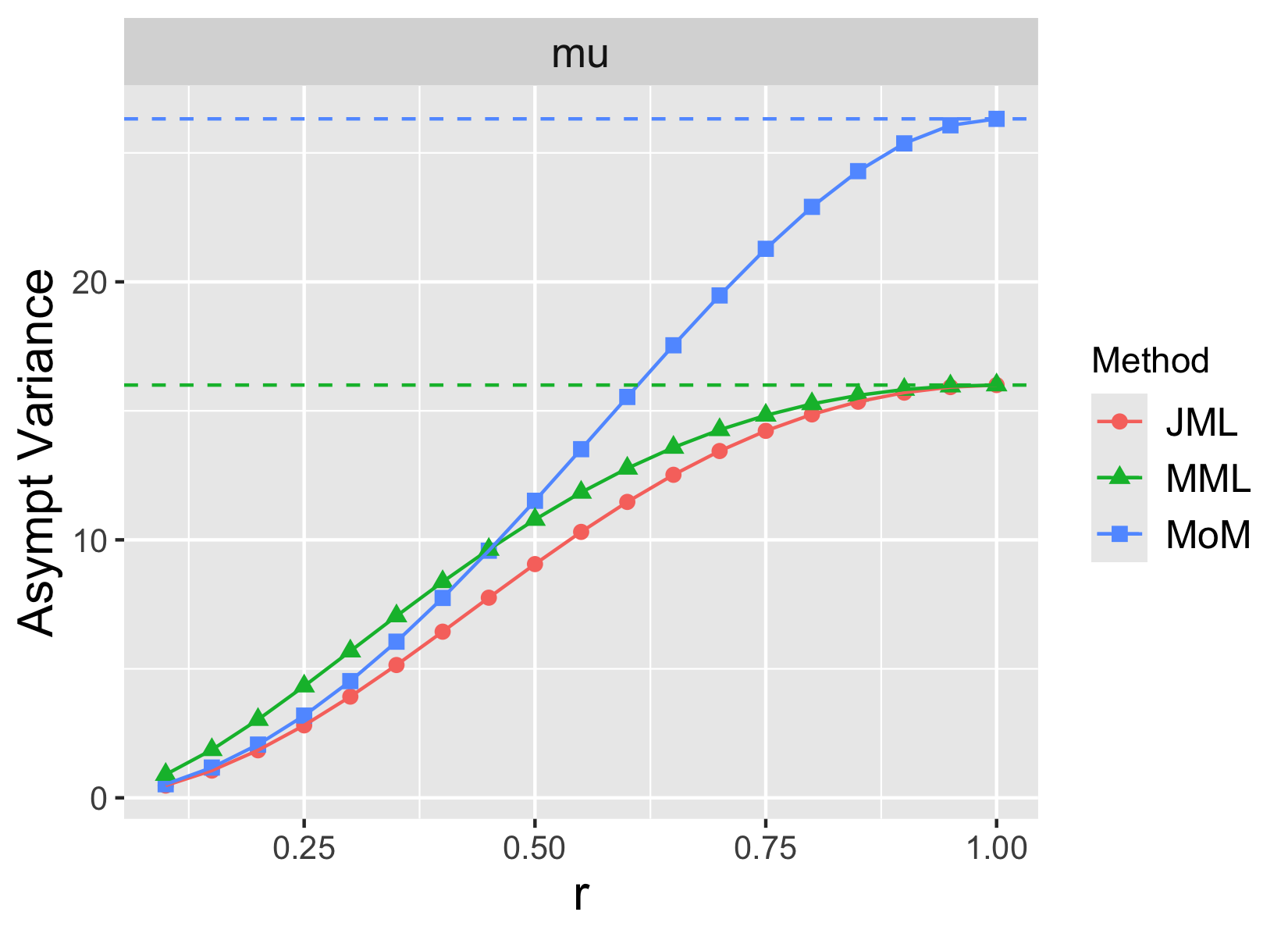}
}
\caption{Asymptotic variances of JML (red, circle), MML (green, triangle), MoM (blue, square), and baseline (dashed) estimators for $\mu_1$ across dependence parameter values $r$ for the bivariate Gumbel distribution.}
\label{f:bvg-variance}
\end{figure}
\Cref{f:bvg-variance} illustrates the (rescaled) asymptotic variances of the estimators from \eqref{e:ex-bvg-base-mom}--\eqref{e:ex-bvg-mml-var} across varying dependence levels $r$, with the optimal coefficients in \eqref{e:bvg-alpha} applied. The dashed horizontal lines indicate the baseline variances for \eqref{e:ex-bvg-base-mom}--\eqref{e:ex-bvg-base-ml}, calculated using moment and ML methods.
The overall trends mirror those observed for $\sigma$ in
Figure~\ref{f:bvg-unknownS} of the main paper.
At $r=1$, representing independence between variables, the moment method exhibits a higher variance of ${\pi^2}\sigma_1^2/6$ compared to the ML method, which has a variance of $\sigma_1^2$. As the dependence strengthens, (i.e., $r$ decreases), the MoM estimator improves substantially. The variance of the MML estimator matches that of the JML estimator in the independent case and both decrease as dependence increases, with the JML estimator achieving a larger reduction.

\section{Regression perspective}
\label{s:discussion:reg}
In Section~\ref{s:ex-bvn} of the main paper, we established that when the joint distribution follows a bivariate Gaussian distribution, the JML estimator for $\tone=(\mu_1, \sigma_1)$ in Proposition~\ref{prop:bvn:joint-mle} coincides with the MML estimator. We aim to revisit this result here from a regression perspective.

Note that if $(\yhifi, \ylofi)$ is modeled by the bivariate Gaussian distribution, the relationship can be reformulated as a simple linear regression model
\begin{align}
\label{e:reg-model}
\yhifi = \alpha + \beta \ylofi + \epsilon,
\end{align}
where $\ylofi\sim \calN(\mu_2,\sigma_2^2) $ is independent of $\epsilon$ and $\epsilon\sim \calN(0,\sigma^2)$. 
In this framework, $\bfgamma = (\alpha, \beta, \sigma^2)$ represents the regression parameter governing the relationship between $\yhifi$ and $\ylofi$, while $\btheta_2 = (\mu_2, \sigma_2)$ denotes the marginal parameters of $\ylofi$. Then, the joint distribution can be written as
\begin{align}
\label{e:reg-joint-pdf}
f_{\yhifi,\ylofi}(y_1,y_2) &= f_{\yhifi|\ylofi}(y_1|y_2, \bfgamma) f_{\ylofi}(y_2, \ttwo)\\
&=f_{\epsilon}(y_1-\alpha -\beta y_2) f_{\ylofi}(y_2,\ttwo),\notag
\end{align}
where $f_{\yhifi,\ylofi}, f_{\yhifi|\ylofi}$, and $f_{\epsilon}$ represent the joint p.d.f.\ of $(\yhifi, \ylofi)$, the conditional p.d.f.\ of $\yhifi$ given $\ylofi$, and the p.d.f.\ of $\epsilon$, respectively.
The likelihood for our observations \eqref{e:data:mf} can then be written as
\begin{align}
\label{e:reg-lik-factorized}
\prod_{i=1}^n f_{\yhifi,\ylofi}(\yhifi_i,\ylofi_i) \prod_{j=1}^m f_{\ylofi}(\ylofi_{n+j}) = \prod_{i=1}^n f_{\yhifi|\ylofi}(\yhifi_i|\ylofi_i) \prod_{j=1}^{n+m} f_{\ylofi}(\ylofi_{j}),
\end{align}
enabling us to find the ML estimator for $\bfgamma$ and $\ttwo$ by separately maximizing
\begin{align}
\label{e:reg-cond}
\prod_{i=1}^n f_{\yhifi|\ylofi}(\yhifi_i|\ylofi_i)
\end{align}
and
\begin{align}
\label{e:reg-marg}
\prod_{j=1}^{n+m} f_{\ylofi}(\ylofi_j).
\end{align}

This factorization streamlines the estimation procedure as follows: First, all marginal observations $\ylofi_1, \dots, \ylofi_{n+m}$ are used to obtain the ML estimator of $\ttwo$ as
\begin{align}
\label{e:reg-ttwo}
    \widehat \mu_2 = \ybart, \quad \widehat\sigma_2^2 = \frac{1}{n+m}\sum_{i=1}^{n+m}(\ylofi_i-\ybart)^2.
\end{align}
Second, the joint observations $(\yhifi_1, \ylofi_1),
\dots, (\yhifi_n, \ylofi_n)$ are utilized to estimate the regression parameter $\bfgamma$.
The well-known result (e.g., \cite{rencher2008linear}, Chapter 6) for the parameters of the linear regression model leads to
\begin{align}
\label{e:reg-alpha}
\widehat{\alpha} &= \ybaro - \widehat{\beta} \ybarto,\\
\label{e:reg-beta}
\widehat{\beta} &= 
\frac{\sum_{1}^n (\ylofi_i-\ybarto)(\yhifi_i-\ybaro)}{\sum_{1}^n (\ylofi_i-\ybarto)^2},\\
\label{e:reg-sig}
\widehat{\sigma}^2 &= \frac{1}{n}\sum_{i=1}^n (\yhifi_i - \widehat Y_i^{(1)})^2 = \frac{1}{n}\left\{\sum_{i=1}^n (\yhifi_i - \ybaro)^2 - \sum_{i=1}^n \widehat{\beta}^2(\ylofi_i - \ybarto)^2 \right\},
\end{align}
where $\widehat Y_i^{(1)} = \ybaro + \widehat{\beta}(\ylofi_i-\ybarto)$.

We can now verify that the obtained estimators agree with Proposition~\ref{prop:bvn:joint-mle} of the main paper.
Under the model \eqref{e:reg-model}, we have
\begin{align}
\label{e:reg-mu1}
\mu_1=\bbE \yhifi &= \bbE(\bbE(\yhifi|\ylofi)) = \alpha + \beta~ \bbE \ylofi,\\
\label{e:reg-sig1}
\sigma_1^2 = \var (\yhifi) &=\var(\bbE(\yhifi|\ylofi)) + \bbE(\var(\yhifi|\ylofi)) = \beta^2 \var (\ylofi) + \sigma^2.
\end{align}
Substituting \eqref{e:reg-ttwo}--\eqref{e:reg-sig} into \eqref{e:reg-mu1}--\eqref{e:reg-sig1} leads to
\begin{align}
\widehat\mu_1 & =  \ybaro+ \widehat{\beta}(\ybart-\ybarto),\\
\widehat\sigma_1^2 &= \frac{1}{n}\sum_{i=1}^n (\yhifi_i - \ybaro)^2 + \widehat{\beta}^2\left\{ \widehat \sigma_2^2 - \frac{1}{n}\sum_{i=1}^{n}(\ylofi_i-\ybarto)^2\right\},\\
\widehat\rho &= \big(\widehat{\beta\frac{\sigma_2}{ \sigma_1}}\big) =  \frac{\widehat \sigma_2}{\widehat \sigma_1}\frac{\sum_{1}^n (\ylofi_i-\ybarto)(\yhifi_i-\ybaro)}{\sum_{1}^n (\ylofi_i-\ybarto)^2},
\end{align}
recovering \eqref{e:bvn-m1}, \eqref{e:bvn-s1} and \eqref{e:bvn-rho} in Proposition~\ref{prop:bvn:joint-mle} of the main paper.

The above observations illustrate when the MFMC estimator could indeed be the optimal estimator of the maximum likelihood criterion with the observations \eqref{e:data:mf}. They also suggest several other interesting points.
First, the factorization in \eqref{e:reg-lik-factorized} relies solely on the independence of $\ylofi$ and $\epsilon$. Thus, the result extends to the case where $\ylofi$ and $\epsilon$ do not follow Gaussian distributions. Provided they remain independent and satisfy the relationship in \eqref{e:reg-model}, we can separately estimate the regression parameters using the joint observations and the distribution of $\ylofi$ using the marginal observations of $\ylofi$. This estimation process remains optimal within this framework.

Second, the MFMC estimator (e.g., as in \eqref{e:mfmc-mom-element}) can be obtained by assuming a linear model between the transformed variables $h(\yjifi)$ and then using estimated least-squares coefficients. In practice, this model may not satisfy the usual regression assumptions. However, applying standard regression techniques, such as data transformations to better satisfy assumptions (e.g., homoscedasticity), can improve the estimator's performance. This approach allows for more efficient estimators by ensuring that the underlying assumptions of the linear regression model are better met.

\section{Importance sampling}
\label{s:discussion:is}
Throughout the main paper, we worked under the assumption that $(\yhifi_i, \ylofi_i), ~ i=1, \dots, n,$ are i.i.d., that is, sampled at random. This means that each vector observation $(\yhifi_i, \ylofi_i)$ carries the same weight $w_i = 1/n$ in the sample. A straightforward extension of our methods is to the case when weights $w_i>0$ are not equal across $i$, as in importance sampling schemes. For example, assuming the case \eqref{e:mfmc-case-m}, the JML estimation would now read as
\begin{align}
\label{e:dis-is-mle}
(\htheta_{1,{\rm jml}}, \htheta_{1,2,{\rm jml}}) = \argmax{\tone, \tot} \sum_{i=1}^n w_i \ell_{(\tone, \ttwo^\ast, \tot)} (\yhifi_i, \ylofi_i).
\end{align}
Similarly, the average $\overline{\bfh(\yjifi)}_n$ in \eqref{e:mom-first}--\eqref{e:mom-second} would become $\sum_{i=1}^n w_i\; \bfh(\yjifi_i),$ assuming the weights are normalized as $\sum_{i=1}^n w_i = 1$.

An example of a multi-fidelity importance sampling scheme is introduced in \cite{kim2025sampling}, where importance weights are directly derived from observed low-fidelity model outputs. Specifically, the weights are calculated as the ratio ${f_{\ylofi}(\ylofi)}/{p_{\ylofi}(\ylofi)}$, with $f_{\ylofi}$ representing the true marginal density of the low-fidelity outputs, and $p_{\ylofi}$ denoting the chosen biasing distribution.
In contrast, other multi-fidelity importance sampling methods focus on the underlying random inputs $x$, where each fidelity output is defined as $Y^{(j)} = g_j(x)$ for some function $g_j$. Here, $x$ may represent a random input used in the forward simulation, such as a PDE parameter. Importance sampling in this context is applied directly within the input space (e.g., \cite{peherstorfer2016multifidelity}, \cite{KRAMER2019}, \cite{pham2022ensemble}).
In both approaches, low-fidelity models are explored first to construct biasing distributions that effectively guide sampling for high-fidelity evaluations.

\bibliography{bibliography.bib}

\end{document}